\documentclass[a4paper,11pt]{article}
\pdfoutput=1 

\usepackage{jheppub} 

\usepackage[T1]{fontenc} 

\usepackage{slashed}
\usepackage{multirow}
\usepackage{enumerate}
\usepackage{enumitem}
\usepackage{float}

\newcommand{\be}{\begin{equation}}
\newcommand{\ee}{\end{equation}}
\newcommand{\bea}{\begin{eqnarray}}
\newcommand{\eea}{\end{eqnarray}}

\newcommand{\met}{\slashed{E}_T}

\newcommand{\iab}{\rm ab^{-1}}
\newcommand{\ifb}{\rm fb^{-1}}
\newcommand{\mltp}{{\mkern-2mu\times\mkern-2mu}}

\title{\boldmath Probing the decoupled seesaw scalar in rare Higgs boson decay}

\author[a]{Yu Gao,}
\author[b,a]{Mingjie Jin,}
\author[c]{Kechen Wang}

\affiliation[a]{Key Laboratory of Particle Astrophysics, Institute of High Energy Physics,\\Chinese Academy of Sciences, Beijing, 100049, China}
\affiliation[b]{Department of Physics, Beijing Normal University, Beijing, 100875, China}
\affiliation[c]{Department of Physics, School of Science, Wuhan University of Technology,\\430070 Wuhan, Hubei, China}

\emailAdd{gaoyu@ihep.ac.cn}
\emailAdd{jinmj@ihep.ac.cn}
\emailAdd{kechen.wang@whut.edu.cn}

\abstract{
The Higgs boson can mix with a singlet scalar that dynamically generates the Majorana mass of the right-handed neutrino $N_R$. 
We show that even a tiny mixing between the Higgs boson and a `decoupled' singlet scalar allows for Higgs-mediated pair production of $N_R$ without significant mixings between the active neutrinos and $N_R$, and thus testable at colliders via a characteristic signal of two same-sign same-flavor lepton pairs, plus missing energy. 
We demonstrate that this search channel is mostly background-free in $pp$-collision and can be a highly sensitive probe of the Higgs-singlet mixing at the current and future $pp$ colliders. Such channel provides a clean signal to discover the singlet scalar and explore the origin of neutrino masses.
}

\begin{document} 
\maketitle
\flushbottom

\section{Introduction}
\label{sec:intro}

The seesaw mechanism has been extensively studied to generate the tiny active neutrino mass. By adding a right-handed neutrino $N_R$ and introducing a Dirac neutrino mass $m_{D}\,\nu_L N_R$, the Standard Model's (SM's) neutrino $\nu_L$ and the hypothetical $N_R$ can mix into the active neutrino $\nu$ and a heavy neutrino $N$, where the active neutrino mass $m_\nu\sim m_D^2/M$ can be brought down to sub-eV scale by having a high-scale Majorana mass $M$ of $N_R$. 

The heavy $N$ is an important target of new physics searches as the crucial seesaw ingredient. At colliders it can be weakly produced via its active neutrino component or by its $N_R$ component's coupling to SM particles. One major difficulty for the weak production is that it is limited by $N_R$'s typically small mixing with the weakly charged $\nu_L$. Nowadays Large Hadron Collider (LHC) searches~\cite{Sirunyan:2018xiv, Aaboud:2018spl} and electroweak precision measurements~\cite{Abada:2013aba, Akhmedov:2013hec} probe the effective $\nu_l-N_R$ mixing parameters $|V_{lN}|^2$~\cite{Atre:2009rg} to $\sim 10^{-5}$. Alternatively, a $N_R$ uncharged under SM gauge interactions can still have a sizable coupling to massive Beyond the Standard Model (BSM) scalars which mix with the SM Higgs boson. Well-motivated scenarios include the $N_R$ coupling to the neutral scalar components in Left-Right symmetric models~\cite{Wyler:1982dd}, gauged $U(1)_{B-L}$ symmetric models~\cite{Mohapatra:1979ia}, and seesaw models with gauge-singlet scalars~\cite{Shoemaker:2010fg}, etc. 

While direct massive BSM mediator production is an important aspect of future high energy searches, their contribution to the rare decay of the SM bosons must be carefully examined for a potentially tiny $N_R$ related mixing. Well known instances include the rare Higgs decay channel $h\rightarrow l^\pm l^\pm 4j$ that is enhanced by the Higgs-BSM scalar mixing~\cite{Shoemaker:2010fg, Maiezza:2015lza, Nemevsek:2016enw}, which have been studied as a good test of lepton number violation (LNV) in case of a Majorana heavy neutrino. 
$N_R$ characteristic and collider-testable rare Higgs decay channels are very important precision searches, especially as so far collider searches show no robust sign of BSM mediators.

Generally, a singlet scalar that dynamically generates the Majorana mass of the $N_R$ can mix with the Higgs boson, which allows the Higgs boson's decay mode of $h \to NN$ without requiring sizeable left-right neutrino mixings. Such a singlet extension of the SM is often motivated for the dynamical $N_R$ mass generation, and it potentially helps understanding several important phenomena: the baryon number asymmetry via a strong first order electroweak phase transition~\cite{Pietroni:1992in,Davies:1996qn}, the abundance of  dark matter~\cite{Silveira:1985rk,McDonald:1993ex} and the issue of vacuum stability~\cite{Clark:2009dc,Lerner:2009xg,Gonderinger:2009jp}, etc. In this study we advocate that such new singlet scalar could be discovered at $pp$ colliders by searching for the fully leptonic rare Higgs decay channel  $h \to NN \to 4l+\met$.

For minimal parameter dependence we start with a {\it small-mixing} scenario of the BSM scalar sector in a Type I seesaw setup where the L-R mixing is typically small,
\bea
{\cal L} &\supset& V(\mathit{\Phi}) + V(S) + \frac{\lambda}{2} |\mathit{\Phi}|^2 S^2 \label{eq:lagrangian}\\
 && + y_N S\bar{N}_R^cN_R + y_D \bar{L}\mathit{\Phi} N_R + c.c. .\nonumber
\eea
$\mathit{\Phi}$ denotes the scalar field(s) that yield the 125 GeV Higgs boson, and $S$ for a singlet scalar field that couples to the right-hand singlet neutrino $N_R$. Here we leave out the specific form of the potentials and only assume `small-mixing', which is a very subleading mixing term $\frac{1}{2} \lambda|\mathit{\Phi}|^2S^2$, so that $V(\mathit{\Phi})$ and $V(S)$ minimize independently. $\mathit{\Phi}$ and $S$ develop their own vacuum expectation values (vev):
$\mathit{\Phi} = (v_{\mathit{\Phi}} + \phi)/\sqrt{2}$ to give rise to electroweak symmetry breaking and generate a Higgs-like boson $\phi$, and $S = v_S +s$ to yield a necessary Majorana mass $m_{N_R}= 2y_N v_S$. The small but non-zero mixing term also portals $N_R$ into $\phi$ decay, yielding  our search signal. 
For simplicity here we neglected the $|H|^2S$ term which has a dimensional coupling, by assuming its coupling is subdominant to $\lambda v_\Phi$.

The $\lambda$ term develops a $\phi-s$ mass mixing term $\lambda v_{\mathit{\Phi}}v_S \cdot \phi s$ after potential minimization, where the scalar mass-square matrix can be parametrized as,
\be 
\begin{array}{c|cc}
& \phi & s \\
\hline 
\phi & m^2_\phi & \lambda v_\phi v_s \\
s & \lambda v_\phi v_s & m_s^2 \\
\end{array}.
\ee
With `small mixing' such that $\lambda v_{\mathit{\Phi}}v_S \ll m_\phi^2, m_s^2$, this matrix is dominated by the diagonal terms and 
the mass eigenstate  \{$h_{1,2}$\}  is a rotation by a small angle $\alpha$ from the gauge eigenstates \{$\phi$,$s$\},
\be 
\left( \begin{array}{c} h_1 \\ h_2\end{array} \right) = 
\left( \begin{array}{cc} 
\cos\alpha & -\sin \alpha \\ 
\sin\alpha & \cos\alpha 
\end{array} 
\right) 
\left( \begin{array}{c}\phi \\ s \end{array} \right).
\ee

Without loss of generality $h_1$ identifies with the SM Higgs boson and contains an $s$-component of the magnitude
\be 
\alpha = \frac{\lambda v_\mathit{\Phi} v_S}{ |m_s^2 -m^2_\phi|},
\label{eq:sin_a}
\ee
which leads to a decay channel $h_1\to N N$ in additional to the SM decays, if kinematically permitted. 
Similarly $h_2$ is mostly $s$ and has only a small $\phi$-component. 
Since $s$ does not couple to the SM fermions, $h_1$'s $s$-component leaves the relative branching ratios unchanged from their SM values.
Such a mixing can be sizable~\cite{Farina:2013fsa,Lopez-Val:2014jva} if $S$ is completely decoupled ($y_N\rightarrow 0$). For small mixings we would take `decoupling' $\mathit{\Phi}, S$ conditions:
\bea 
\sin^2\alpha & \ll & 1, \\
\lambda\cdot {\rm max}(v_S^2, v_\mathit{\Phi}^2) & \ll & {\rm min}(m_s^2, m_\phi^2).
\label{eq:decoupling_limit}
\eea
The latter simplifies to $\lambda v_S^2\le {\cal O}(10^{-2})m_{h_1}^2$ if we consider an $h_2$ comparable to or heavier than the Higgs boson. 

A finite $y_N$ leads to the $h_1$ decay partial width
\be 
\Gamma _{h_1\to N N} = \frac{1}{2}\sin^2\alpha\cdot\frac{y_N^2m_{h_1}}{8\pi}\left(1-\frac{4m_N^2}{m_{h_1}^2} \right)^{3/2}
\ee
for each $N$ species, where $m_{h_1}$=125 GeV, $m_N =2 y_N v_S$ and the factor $1/2$ is due to $N$ being Majorana. 
Given the rather small SM Higgs boson's decay width $\Gamma_{h}\sim$ 4 MeV and relatively light heavy neutrino mass $m_{N} \lesssim m_{h_1}/2$, the branching ratio of $h_1$ decay to $NN$ is
\be 
{\rm BR}_{h_1\to N N} = \frac{\Gamma_{h_1\to N N}}{\Gamma_{h}+\Gamma_{h_1\to N N}}.
\label{eq:BF_NN}
\ee
This leads to ${\rm BR}_{h_1\to N N} \approx \Gamma_{h_1\to N N}/\Gamma_{h}$ for $\sin^2\alpha \ll 1$. In both Majorana and Dirac $N_R$ cases, the subsequent $NN$ decay can lead to a characteristic two same-sign, same-flavor lepton pair, plus missing transverse energy signal $e^\pm e^\pm \mu^\mp\mu^\mp + \met$.

\section{Signal at $pp$ colliders}
\label{sec:sig}

\begin{figure}[h]
\centering
\includegraphics[width=15cm,height=5cm]{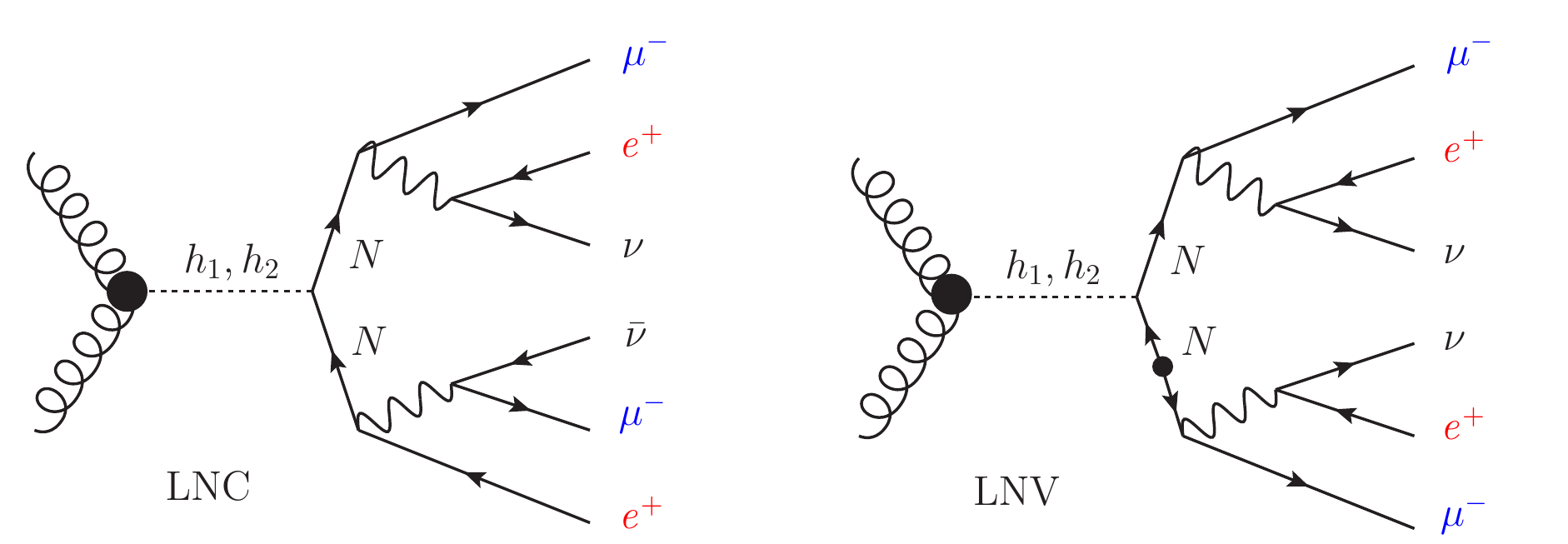}
\caption{Representative $NN$ processes that contain two pairs of leptons of the same-sign and the same-flavor. The LNC diagram (left) requires $N$ couples to at least two lepton flavors.}
\label{fig:feynman}
\end{figure}

The Higgs-like $h_1$ can be copiously produced via gluon fusion during $pp$ collisions with a cross-section $\sigma_{h_1} = \sigma_{pp\to h}$.
$NN$ decay then yields a characteristic and {\it remarkably clean} signal of two same-sign, same-flavor (SSSF) lepton pairs as illustrated in Fig.~\ref{fig:feynman}. 
The two lepton pairs should have different lepton flavors and opposite charges (i.e. $e^\pm e^\pm \mu^\mp \mu^\mp$ and no opposite-sign, same-flavor lepton pair exists).
This channel differs from the LNV $NN\to l^\pm l^\pm 4j$ case that it does not necessarily violate lepton number conservation (LNC) if $N$ couples to more than one lepton flavor, as the final state $\nu$ can be either neutrinos or anti-neutrinos.  
Therefore, this final state can be present either $N$ is Majorana or Dirac. 

With a Majorana $N$, the $NN$ system has a higher decay branching to this final state as one $N$ can also decay as its own antiparticle.
In such case, for each species of $N$ the branching factor of the signal final state from $NN$ is obtained by summing up the LNV and LNC processes,
\bea
{\rm BF}_{\rm sig.} 
&\equiv& {\rm BF}_{\rm sig.}(NN \to e^\pm e^\pm \mu^\mp \mu^\mp  2\nu) \nonumber \\
&=& \frac{1}{2} \sum_{i\neq j} \left( \frac{|V_{iN}|^2}{\Sigma^2} {\rm BF}^l_j \right)^2 \ \ {\rm (for\ LNV)}  \label{eq:BF_NNsig} \\
  &+&  \frac{1}{2} \sum_{i\neq j} \left( \frac{|V_{iN}|^2}{\Sigma^2} {\rm BF}^l_j \right)\left( \frac{|V_{jN}|^2}{\Sigma^2} {\rm BF}^l_i \right) \ \ {\rm (for\ LNC),} \nonumber
\eea
where $\Sigma^2 \equiv |V_{eN}|^2+|V_{\mu N}|^2+|V_{\tau N}|^2$, and ${i,j}=\{e,\mu \}$ for lepton flavors.
${\rm BF}^l_i$ is the $W^{(*)} \rightarrow l_i \nu$ branching fraction in $N$ decay. 
Note the $N\to ll'\nu$ requires charged weak gauge boson mediation, which makes ${\rm BF}^l_i$ depend on both $m_N$ and whether BSM mediators are present in $N$ decay. The maximal BF$_{\rm sig.}$ is reached if $N$ only couples to one lepton flavor, and ${\rm BF}_{\rm sig.} \leq ({\rm BF}^l)^2/2$.

Similarly, the $h_{2}$ scalar can also be produced via its $\phi$-component at a cross-section $\sigma_{h_2} = \sigma_{pp\to h_2} = \left.\sin^2\alpha \cdot \sigma_{pp\to h}\right|_{m^{\rm kin.}_h=m_{h_2}}$,
where the latter is the SM Higgs-like boson production cross-section evaluated for a different kinematical mass $m_{h_2}$.
Since both the $h_1\rightarrow NN$ branching and the $h_2$'s production cross-section scale as $\sin^2\alpha$, assuming a near-unity branching of $h_2\rightarrow NN$, their contributions are the same order in terms of $h-s$ mixing. 

We note that $h_2$ decay modes depend on the relative size between $m_{h_1}, m_{h_2}$ and $m_N$. 
When $m_{h_2} >2 m_{h_1}$, the $h_2\to h_1 h_1$ channel opens up, but it is suppressed at small $\lambda$. 
$h_2$ may also decay to the SM fermion/gauge boson pairs, with an $\alpha^2$ suppression. 
Therefore, in case of a kinematically accessible $m_{h_2} >2  m_N$ and a small $\lambda$, $h_2\to NN$ dominates $h_2$ decays and acquire a near-unity branching.
In case $m_{h_2} \ll m_{h_1}$ and $m_{h_2} < 2 m_N$, however, the relative branching ratios become less transparent as $h_2$ decays via the $\phi$- and $s$- components compete between the $\alpha^2$-dependent two-body channels and the $|V_{lN}|^2$, $m_N$-dependent multi-body channels. 

The total signal cross-section is then
\begin{eqnarray}
\sigma_{\rm sig.} = & & \sigma_{h_1} \cdot {\rm BR}_{h_1\to NN} \cdot {\rm BF}_{\rm sig.} \cdot A_{\rm eff}^{\rm (1)} \nonumber \\
                    &+& \sigma_{h_2} \cdot {\rm BR}_{h_2\to NN} \cdot {\rm BF}_{\rm sig.} \cdot A_{\rm eff}^{\rm (2)} \,\,,
\label{eq:xsec}
\end{eqnarray}
where $A_{\rm eff}^{\rm (1,2)}$ is the detector selection efficiency for $h_1,h_2$ mediated event samples, which can be separately studied via Monte Carlo simulations. $h_2$'s event rate may quickly become subdominant with a heavy $h_2$ mass, and with $m_{h_2}\sim {\cal O}(m_{h_1})$ it can still be the main contributor due to its near-unity decay branching into $NN$. 
Note this $h_{1,2}$ mediated signal rate bypasses the vertex suppression from $\nu_L -N_R$ mixing, and serves as a probe for low $\nu_L-N_R$ mixing scenarios, e.g. Type-I seesaw models.

For convenience in phenomenology study we consider the simplified case where only the mass of one heavy neutrino $N_1$ lies below $\frac{1}{2} m_{h_1}$ and $N_1$ mixes with only one active neutrino flavor $\nu_{\mu}$. 
In this case only the LNV diagram in Fig.~\ref{fig:feynman} contributes, and ${\rm BF}_{\rm sig.}$ reaches its maximum ${\rm BF}_{\rm sig.}=0.16\%$. 
Complication rises when both $m_N$ and $|V_{lN}|^2$ are small the $N$ lifetime can be quite long, which leads to displaced decay at colliders. Our study relies on reconstructing decay kinematics so we require $N$ decays inside the detector. Note that this decay-length issue can be circumvented in BSM models: $|V_{lN}|^2$ can be enhanced by inverse-seesaw~\cite{Mohapatra:1986aw} after introducing mixture into an extra singlet fermion; also if $N_R$ is charged under BSM interactions, e.g. in extra-$U(1)$ and Left-Right symmetric models, $N$ gains extra partial width that helps make its decay visible but it also suppresses BF$^{l}$ at the same time.

The major kinematic difference from a similar large invariant-mass search for doubly-charged Higgs~\cite{Du:2018eaw,Aaboud:2017qph} is that our four leptons are much softer in energy, only adding up to a fraction of the Higgs boson mass in the optimal mass range $m_N< \frac{1}{2} m_{h_1}$. 
In addition, the same-sign lepton pair must rise from different parent particles in the $NN$ decay, leading to different kinematic topologies.
Moreover, our final state contain two neutrinos which act as missing energy.

\section{Background and multivariate analysis}
\label{sec:bkg}

The $e^\pm e^\pm \mu^\mp \mu^\mp +\met$ signal has very few SM background events at the $pp$ colliders due to the presence of two pairs of SSSF leptons. 
The irreducible background includes contribution from $pp\to 4\tau, WWZ(Z\to2\tau)$ channels in which the flavor combinatorics among the four $e/\mu$ leptons after $\tau, W$ decays can produce the same final state.
Because $4\tau$ channel is mainly $ZZ \to 4\tau$ after all selection cuts, it is related to the di-boson production, and is the leading background at lower invariant masses.
The $pp\to 4W$ channel is much subdominant due to its higher number of weak interaction vertices.
Another potential background may arise from events with jets, usually from decays of hadrons into leptons, misidentified as leptons. 
However, a reliable fake-lepton background is usually obtained by the ``data-driven'' approach, which relies on experimental data. A very relevant experiment study~\cite{Aaboud:2017qph} which has a similar signal final state shows that the background from doubly faked leptons can be subdominant to di-boson production. Besides, our signal requires two muons while faked leptons are mostly electrons. Thus, it might indicate that the fake-lepton contribution is controllable in our case.
We include $4\tau$ and $WWZ$ channels as the main background in our $pp$ collision event simulations, and leave the fake-lepton process for further study when data become available.

We quote the measured Higgs boson's single production cross-section via gluon fusion process $\sigma_{pp\to h}=54.7$ pb at 14-TeV ~\cite{gghCrs14TeV} and 740 pb at 100-TeV~\cite{gghCrs100TeV}.
As $h_2$ is also mostly produced by gluon fusion, we adopt the N$^3$LO scalar mass dependence~\cite{Anastasiou:2016hlm} to scale the ${h_2}$ production rate in $\sigma_{h_2}$.
For data simulations we use MadGraph5\_aMC@NLO version 2.6.1~\cite{Alwall:2014hca} for parton level event generation, Pythia6~\cite{Sjostrand:2006za} for the parton shower and hadronization, and Delphes 3.4.1~\cite{deFavereau:2013fsa} for detector simulation.

At the HL-LHC (FCC-hh/SppC), we generate about 150 (165) million $4\tau$,  10 (15) million $WWZ$, and 0.1 million signal events for kinematic analyses.
Simulated events are analyzed to reject the SM background and maximize the signal significance. First the following pre-selection is applied:
(a) exactly four leptons with $p_{\rm T}(l_{1,2})>$ 10 GeV and $p_{\rm T}(l_{3,4})>$ 5 GeV;
(b) one same-sign $e^\pm e^\pm$ pair and one $\mu^\mp \mu^\mp$ pair;
(c) no tagged $\tau$ leptons or $b$-jets.

After pre-selection, in order to construct useful kinematical observables, the stransverse mass $M_{\rm T2}$\cite{Cheng:2008hk,Lester:1999tx,Barr:2003rg} is exploited to identify the electron and muon from the same heavy neutrino decay. 
Since the final state has four leptons (e.g. $\mu_1^{+}\mu_2^{+}e_1^{-}e_2^{-}$) and the two leptons from the same $N$ decay must have opposite sign and different flavor (i.e. $\mu^+e^-$), we calculate the $M_{\rm T2}$ values corresponding to two different combinations of the two-lepton system for $\mu_1$ (i.e. $\mu_1^{+}e_1^{-}$ and $\mu_1^{+}e_2^{-}$). The combination with smaller $M_{\rm T2}$ value (e.g. $\mu_1^{+}e_2^{-}$) is considered from the same $N$ decay, while the remaining lepton pair (i.e. $\mu_2^{+}e_1^{-}$) is considered to be from the other $N$. 
We then input a comprehensive collection of 53 kinematic observables to the TMVA package~\cite{Hocker:2007ht} to perform the MVA.
\begin{enumerate}[label*=\Roman*.]

\item global observables:
\begin{enumerate}[label*=\arabic*)]
\item the missing transverse energy $\met$;
\item $M_{\rm T2}(\mu_{1}+e,\mu_{2}+e)$.
\end{enumerate}

\item observables for the system of lepton(s):
\begin{enumerate}[label*=\arabic*)]
\item $p_{\rm T}$ and the pseudorapidity $\eta$ of lepton $p_{\rm T}(\mu_{1})$, $p_{\rm T}(\mu_{2})$, $p_{\rm T}(e_{1})$, $p_{\rm T}(e_{2})$, $\eta(\mu_{1})$, $\eta(\mu_{2})$, $\eta(e_{1})$, $\eta(e_{2})$;
\item $p_{\rm T}$, $\eta$ and the invariant mass $M$ of leptons $p_{\rm T}(\mu_{1}+\mu_{2})$, $\eta(\mu_{1}+\mu_{2})$, $M(\mu_{1}+\mu_{2})$, $p_{\rm T}(e_{1}+e_{2})$, $\eta(e_{1}+e_{2})$, $M(e_{1}+e_{2})$, $p_{\rm T}(\mu_{1}+e)$, $\eta(\mu_{1}+e)$, $M(\mu_{1}+e)$, $p_{\rm T}(\mu_{2}+e)$, $\eta(\mu_{2}+e)$, $M(\mu_{2}+e)$, $p_{\rm T}(\mu_{1}+\mu_{2}+e_{1}+e_{2})$, $\eta(\mu_{1}+\mu_{2}+e_{1}+e_{2})$, $M(\mu_{1}+\mu_{2}+e_{1}+e_{2})$;
\item the pseudorapidity difference $\Delta\eta$ and the azimuthal angle difference $\Delta\phi$ of the system of leptons $\Delta\eta(\mu_{1},\mu_{2})$, $\Delta\phi(\mu_{1},\mu_{2})$, $\Delta\eta(e_{1},e_{2})$, $\Delta\phi(e_{1},e_{2})$, $\Delta\eta(\mu_{1},e)$, $\Delta\phi(\mu_{1},e)$, $\Delta\eta(\mu_{2},e)$, $\Delta\phi(\mu_{2},e)$, $\Delta\eta(\mu_{1}+e,\mu_{2}+e)$, $\Delta\phi(\mu_{1}+e,\mu_{2}+e)$.
\end{enumerate}

\item observables between the missing energy and lepton(s):
\begin{enumerate}[label*=\arabic*)]
\item $\Delta\phi$ between the missing transverse momentum and lepton(s) 
$\Delta\phi(\met, \mu_{1})$, $\Delta\phi(\met, \mu_{2})$, $\Delta\phi(\met, e_{1})$, $\Delta\phi(\met, e_{2})$, $\Delta\phi(\met, \mu_{1}+\mu_{2})$, $\Delta\phi(\met, e_{1}+e_{2})$, $\Delta\phi(\met, \mu_{1}+e)$, $\Delta\phi(\met, \mu_{2}+e)$, $\Delta\phi(\met, \mu_{1}+\mu_{2}+e_{1}+e_{2})$;
\item the transverse mass $M_{\rm T}$ of the system formed by the missing momentum plus lepton(s) $M_{\rm T}(\met, \mu_{1})$, $M_{\rm T}(\met, \mu_{2})$, $M_{\rm T}(\met, e_{1})$, $M_{\rm T}(\met, e_{2})$, $M_{\rm T}(\met, \mu_{1}+\mu_{2})$, $M_{\rm T}(\met, e_{1}+e_{2})$, $M_{\rm T}(\met, \mu_{1}+e)$, $M_{\rm T}(\met, \mu_{2}+e)$, $M_{\rm T}(\met, \mu_{1}+\mu_{2}+e_{1}+e_{2})$.
\end{enumerate}

\end{enumerate}
Here $\mu_1$, $\mu_2$, $e_1$, $e_2$ denote muons or electrons with the first or second leading transverse momentum, while for ($\mu_1+e$) or $(\mu_1, e)$ the muon and electron are two leptons from the same N decay.
The variables can be ranked according to their importance in the TMVA analysis. For the benchmark case with signal from $h_1 \to NN$ and $m_N = 20$ GeV,  at the HL-LHC, the most useful observables are found to be
$M_{\rm T2}(\mu_{1}+e,\mu_{2}+e)$, $\Delta\phi(\mu_{2},e)$, $M(\mu_{1}+e)$, $\Delta\phi(\mu_{1},e)$, $M(\mu_{2}+e)$, $\Delta\eta(\mu_{2},e)$, $\Delta\eta(\mu_{1},e)$, $M_{\rm T}(\met, \mu_{1}+\mu_{2}+e_{1}+e_{2})$, $\Delta\phi(\met, \mu_{1}+\mu_{2})$, $\Delta\phi(\met, e_{1}+e_{2})$, $\Delta\phi(\mu_{1},\mu_{2})$, $M(\mu_{1}+\mu_{2}+e_{1}+e_{2})$.
We note that since the kinematics vary with $m_N$, the rank may also change for different heavy neutrino masses.

Fig.~\ref{fig:ObsmN20} in the Appendix~\ref{appendix:distribution} illustrates the distributions of a few input observables for a $m_N$=20 GeV signal sample (black, filled), and those for the SM background including $4\tau$ (red) and $WWZ$ (blue) processes after the pre-selection cuts. Many observables show clear distinctions between the signal and background.
It is worth noting that the signal's transverse mass $M_{\rm T2}$ distribution develops a sharp end-point around $m_N$, which means that it can be used to determine the mass of the heavy neutrino as well if this new physics scenario is discovered.

\begin{figure}[h]
\centering
\includegraphics[width=11cm,height=7cm]{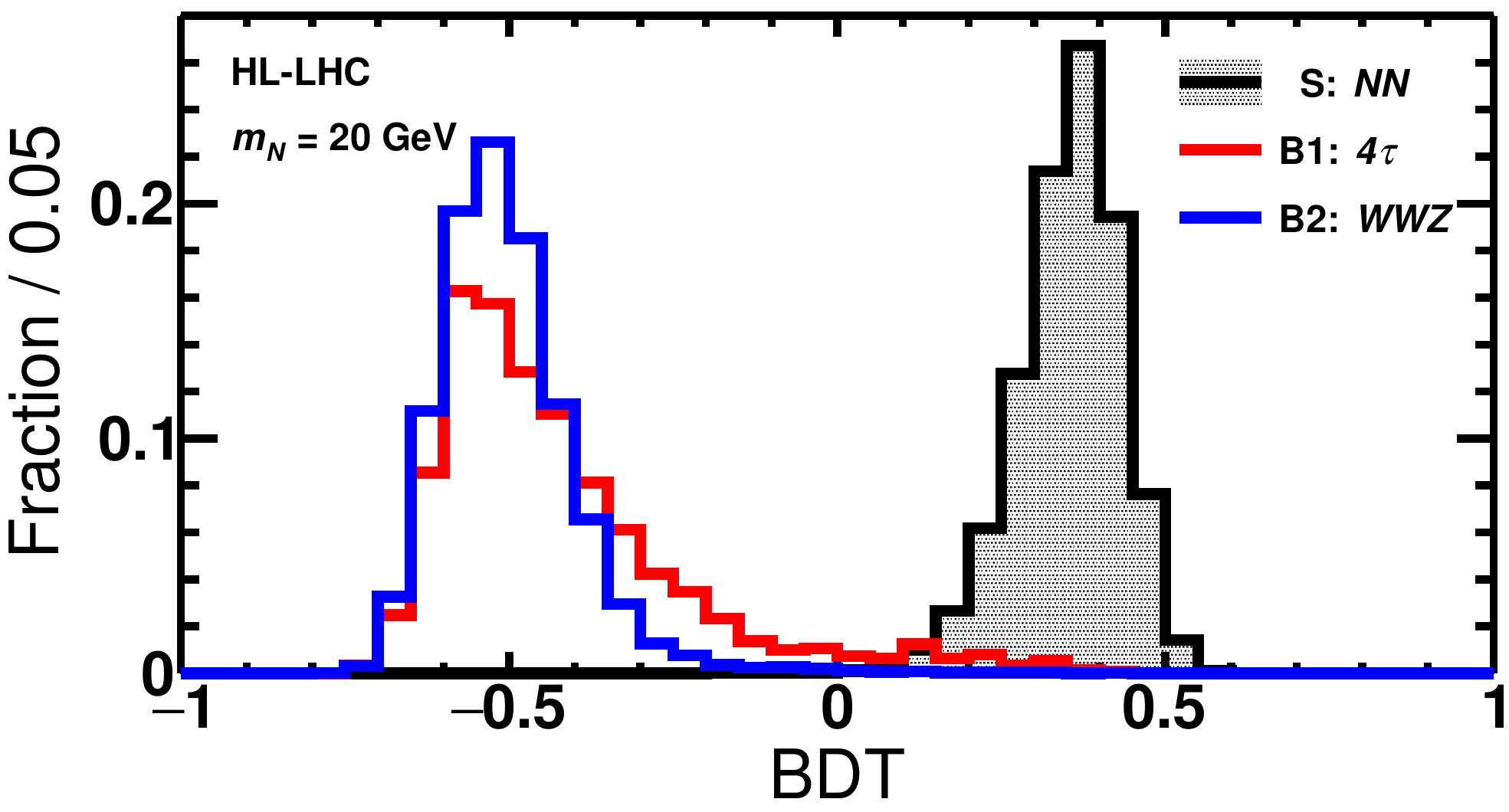}  
\caption{Distributions of BDT responses for the signal event sample (black, filled), and those for the SM background samples including $4\tau$ (red) and $WWZ$ (blue).
This figure illustrates an $h_1$ decay signal by gluon fusion at the HL-LHC, with $m_N$=20 GeV.}
\label{fig:BDTmN20}
\end{figure}

The resulting Boosted Decision Trees (BDT) distributions from the MVA demonstrate excellent separation between signal and background, as shown in Fig.~\ref{fig:BDTmN20}, which can reduce the background to a negligible level. Since the kinematics vary with $m_N$, here we show a representative result of $h_1$ decay with $m_N=20$ GeV. Table~\ref{tab:cutFlow14TeV} lists the corresponding background event rate estimated after pre-selection and the MVA, where the BDT cut is optimized for the signal significance, given by
\be 
\sigma_{\rm stat} = \sqrt{2 [(N_s+N_b) {\rm ln}(1+\frac{N_s}{N_b}) - N_s ] },
\label{eqn:statSgf}
\ee
where $N_s$ ($N_b$) is the number of signal (total background) events after all selection cuts with 3 $\iab$ integrated luminosity at the HL-LHC or 20 $\iab$ integrated luminosity at the FCC-hh/SppC.
Note that Table~\ref{tab:cutFlow14TeV} does not include $h_2$ contribution; $h_2$ increases the signal but the BDT cut needs to be re-optimized according to its fraction in the signal.\\

\begin{table}[h]
\centering
\begin{tabular}{|c|c|ccc|}
\hline
\multicolumn{2}{|c|}{}                          & $NN$  & $4\tau$                  & $WWZ$ \\
\hline  
\multirow{3}{*}{HL-LHC}      & initial     & 16.4   & $2.9\times 10^5$   & $1.1\times 10^3$ \\
                             & pre-selection       &   1.8  & 7.0                            & 1.4 \\
                             & BDT $>$ 0.267  &   1.5   & $6.6\times 10^{-2}$ & $9.6 \times 10^{-4}$ \\
\hline
\multirow{3}{*}{FCC-hh/SppC} & initial  & 1481 & $1.2\times 10^7$    & $1.1\times 10^5$ \\
                             & pre-selection        &   343 & 1183                        & 346 \\
                             & BDT $>$ 0.170   &   298  & 7.5                           & $1.5 \times 10^{-1}$ \\
\hline
\end{tabular}
\caption{Event rates for the signal from $h_1\to NN \to e^\pm e^\pm \mu^\mp \mu^\mp  2\nu$ with $m_N=20$ GeV and ${\rm BF}_{h_1\to e^\pm e^\pm \mu^\mp \mu^\mp  2\nu}=10^{-7}$ and for background processes of $4\tau$ and $WWZ$. The numbers at the HL-LHC (FCC-hh/SppC) correspond to center-of-mass energy $\sqrt{s}=14 (100)$ TeV and 3 (20) $\iab$ integrated luminosity. 
}
\label{tab:cutFlow14TeV}
\end{table}

\section{Collider sensitivites}
\label{sec:result}

Since the final state kinematics vary with the heavy neutrino mass, We simulate the signal sample and perform the MVA for each $M_N$ case from 10 to 60 GeV, at increment of 10 GeV.
The corresponding BDT distributions at the HL-LHC with $\sqrt{s} = 14$ TeV are shown in Fig.~\ref{fig:BDT14TeV}, and distributions at the FCC-hh/SppC with $\sqrt{s} = 100$ TeV are shown in Fig.~\ref{fig:BDT100TeV}, see Appendix~\ref{appendix:BDTdistribution}.

As $m_N$ increases, the kinematics of the $4 \tau$ background becomes similar to that of the signal process $NN$, which renders its BDT distribution largely overlapping with that of the signal.
At each $m_N$, the BDT criterion is optimized for maximal signal significance from its BDT distribution.
The selection efficiencies for both $NN$ signal and background processes at the HL-LHC and FCC-hh/SppC for different heavy neutrino masses are presented in Table~\ref{tab:allEfficiencies}.

\begin{figure}[h]
\centering
\includegraphics[width=11cm,height=7cm]{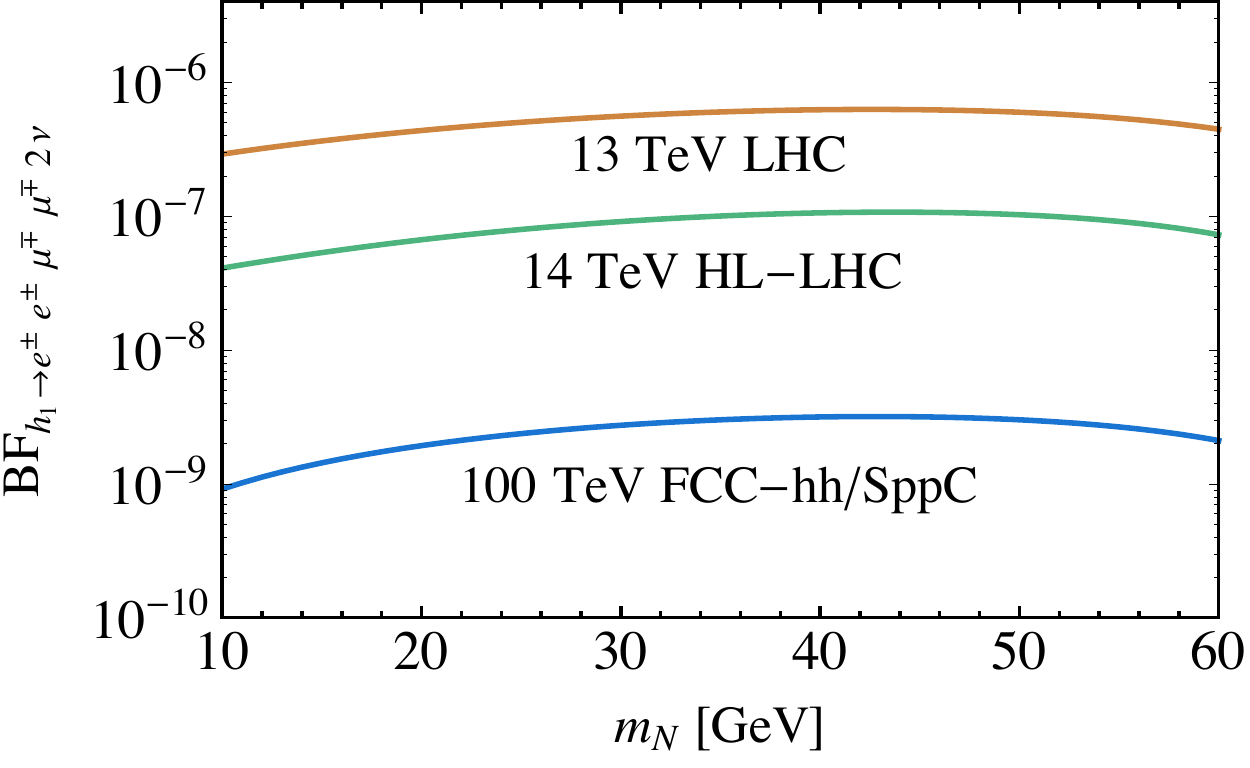}  
\caption{95\% C.L. limits on the combined branching fraction 
${\rm BF}_{h_1\to e^\pm e^\pm \mu^\mp \mu^\mp  2\nu}={\rm BF}_{h_1\to NN} \cdot {\rm BF}_{\rm sig.}$ 
at the 13-TeV LHC, 14-TeV HL-LHC and 100-TeV FCC-hh/SppC with 280 $\ifb$, 3 $\iab$, 20 $\iab$ luminosities, respectively.
For these limits, $h_2$ is assumed to be heavy and only $h_1$ decay contributes to the $NN$ signal.
}
\label{fig:limits}
\end{figure}

Fig.~\ref{fig:limits} shows the 95\% confidence level (C.L.) limits on the combined branching fraction of Higgs boson into our final state when $h_2$ is heavy and only $h_1$ is relevant to the $NN$ signal.
Here, the 95\% C.L. limits correspond to 2-$\sigma$ signal significance given by the Eq.~(\ref{eqn:statSgf}).
The number of events after all selections, $N_s (N_b)$, is the product of the production cross-section, luminosity and total selection efficiency
for the signal (total background). Note the signal production cross-section is proportional to the Higgs decay branching fraction, cf. Eq.~(\ref{eq:xsec}).
The $m_N$ dependence of the limit is the result of two competing reasons: the four-lepton pre-selection efficiencies increase at larger $m_N$, while the BDT cuts reject background more efficiently at lower $m_N$.
When $m_N = 20$ GeV, the combined branching fraction can be probed to as tiny as $4.4 \times 10^{-7}$, $6.7 \times 10^{-8}$, $1.9 \times 10^{-9}$ at the 13, 14, 100 TeV $pp$ collision runs, respectively.
Since the kinematics are quite similar between 13-TeV and 14-TeV, the 13-TeV limits are derived by adopting the same detector selection efficiencies as those for 14-TeV. $\sigma_{pp\to h}=48.6$ pb at 13-TeV~[23] is used for signal production cross-section. For background processes, their production cross-sections at 13~TeV are calculated by MadGraph5.

\begin{figure}[h]   
\centering
\includegraphics[width=11cm,height=7cm]{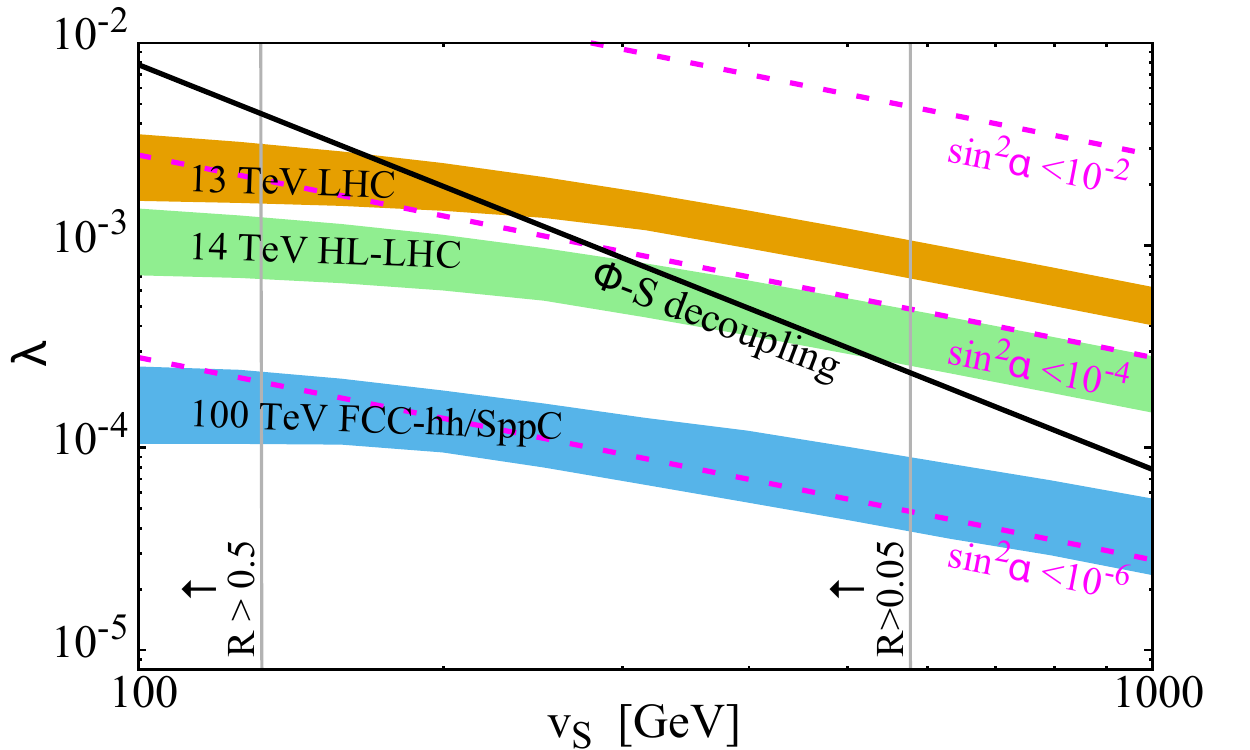}   
\caption{Projected 95\% C.L. sensitivity on $v_S$ and $\lambda$ for a benchmark $m_{h_2}$=150 GeV scenario. The yellow, green and blue bands denote limits at the current 13-TeV LHC, 14-TeV HL-LHC and 100-TeV FCC-hh/SppC with  280 $\ifb$, 3 $\iab$, 20 $\iab$ luminosity, respectively.
The black solid `$\mathit{\Phi}-S$ decoupling' line represents decoupled $\mathit{\Phi}, S$ sectors, satisfying $\lambda v_S^2 \le 10^{-2}m_{h_1}^2$.
}
\label{fig:sensitivity}
\end{figure}

Now we interpret these limits in the content of our `small mixing' scenario. By Eq.~(\ref{eq:sin_a}) the mixing angle is most sensitive to the mass-square difference between $\phi$ and $s$, which can be approximated the mass-square difference between $h_1$ and $h_2$, off by a correction $\Delta m^2_{s,\phi}=\Delta m^2_{h_2,h_1}\cdot[1+{\cal O}(\sin^2\alpha)]$. Generally, $m_s$ can be lighter than $v_S$ and it is possible to have $h_2$ at the same (or lower) mass scale as the SM Higgs, so that $|\Delta m^2_{s,\phi}|\sim {m_h^2}$ even if $v_S$ is more massive, at the TeV scale or above. 
So we adopt a 150 GeV $h_2$ as a showcase scenario, with the consideration that
an $m_{h_2}$ below the $2m_W$ threshold can avoid a dramatic increment in ${\rm BF}_{h_2\to WW,ZZ}$ which suppresses the $NN$ branching and complicate our physics picture.
In this scenario, both $h_2$ and $h_1$ decays contribute significantly. Fig.~\ref{fig:sensitivity} illustrates the 95\% C.L. contour limits on $\lambda, v_S$ from various $pp$ collision runs. 
The width of each filled band is due to varying $m_N$ from 10 to 60 GeV, which affects the cut efficiencies. $R$ is the fraction of $h_1$ contribution in the total $NN$ signal events.
$h_1$ contributes mostly to the signal at low $v_S$, while $h_2$ can also contribute due to its relatively light mass, and becomes dominant towards $v_S\sim$ TeV. 
The limits indicate that the 13-TeV LHC with 280 fb$^{-1}$ data can access the weak-scale $v_S$ part of the decoupling region, while the future HL-LHC and FCC-hh/SppC can be sensitive to even lower $\mathit{\Phi}-S$ mixing region, 
where the singlet components in $h_1$ can be as low as $10^{-4}$ to $10^{-6}$.

For heavier $h_2$, at the TeV scale, $h_2$'s contribution to the signal diminishes and the mixing $\sin^2\alpha$ is suppressed by the TeV $m_s$. However, since $y_N=m_N/(2v_S)$, an $m_N<\frac{1}{2}m_{h1}$ also significantly suppresses the $hNN$ coupling, leading to $m_{s}^{-4}$ suppression in the signal rate. 
Therefore, a large $m_s$ would require a sizable $\lambda$ to yield a detectable signal rate, which could violate the decoupling condition as $v_S$ is generally expected to be comparable to or higher than $m_s$.
In such cases, the $h-s$ mixing needs to be parametrized under specific models where $\mathit{\Phi},S$ sectors are no longer independent. Note in $SU(2)_L$ doublet(s)+singlet scalar models, the joint scalar potential minimization removes $\lambda$ dependence from the diagonal scalar mass-square terms at the tree-level, indicating Eq.~(\ref{eq:decoupling_limit}) to be a strong requirement if $\lambda\ll 1$, which suppresses the singlet's loop contribution to the Higgs mass.

\section{Conclusion}
\label{sec:discussion}

We show that the fully leptonic Higgs decay channel $e^\pm e^\pm \mu^\mp\mu^\mp + \met$ is almost SM background-free and is accessible in both the lepton number violating and conserving heavy neutrino $N$ decays. This rare channel can be a high-precision probe for the mixing between the Higgs boson and a singlet scalar boson which dynamically gives the $N$ mass in the $|V_{lN}|$-suppressed simple seesaw scheme.
We consider a generic `decoupled' scenario where the Higgs and singlet scalar bosons only share a small mixing, and demonstrate that the signal channel is capable of testing such decoupled scenario if the singlet and the Higgs bosons are in the comparable mass range.
Comprehensive signal and SM background analyses with machine-learning MVA yield stringent limits on the signal Higgs boson decay branching fraction at the current and future $pp$ colliders. For a benchmark case with the singlet-dominated $h_2$ at 150 GeV, the 13-TeV LHC, future HL-LHC and FCC-hh/SppC are sensitive to the Higgs-singlet mixing to $10^{-3}$, $10^{-4}$ and $10^{-6}$ with 280 $\ifb$, 3 $\iab$, and 20 $\iab$ $pp$ collision luminosities, respectively. In addition, this test does not require a sizable left-right neutrino mixing and it is complimentary to the existing $|V_{lN}|^2$-based searches.

\appendix

\newpage
\section{Distributions of representative input observables}
\label{appendix:distribution}

In Fig.~\ref{fig:ObsmN20}, we show the distributions of some observables for the signal (black, filled), and those for the SM background including $4\tau$ (red) and $WWZ$ (blue) processes after the pre-selection cuts. 
These plots illustrate a $h_1$ decay signal by gluon fusion at the HL-LHC, with $m_N$=20 GeV.

\begin{figure}[h]
\centering
\includegraphics[width=4.8cm,height=3cm]{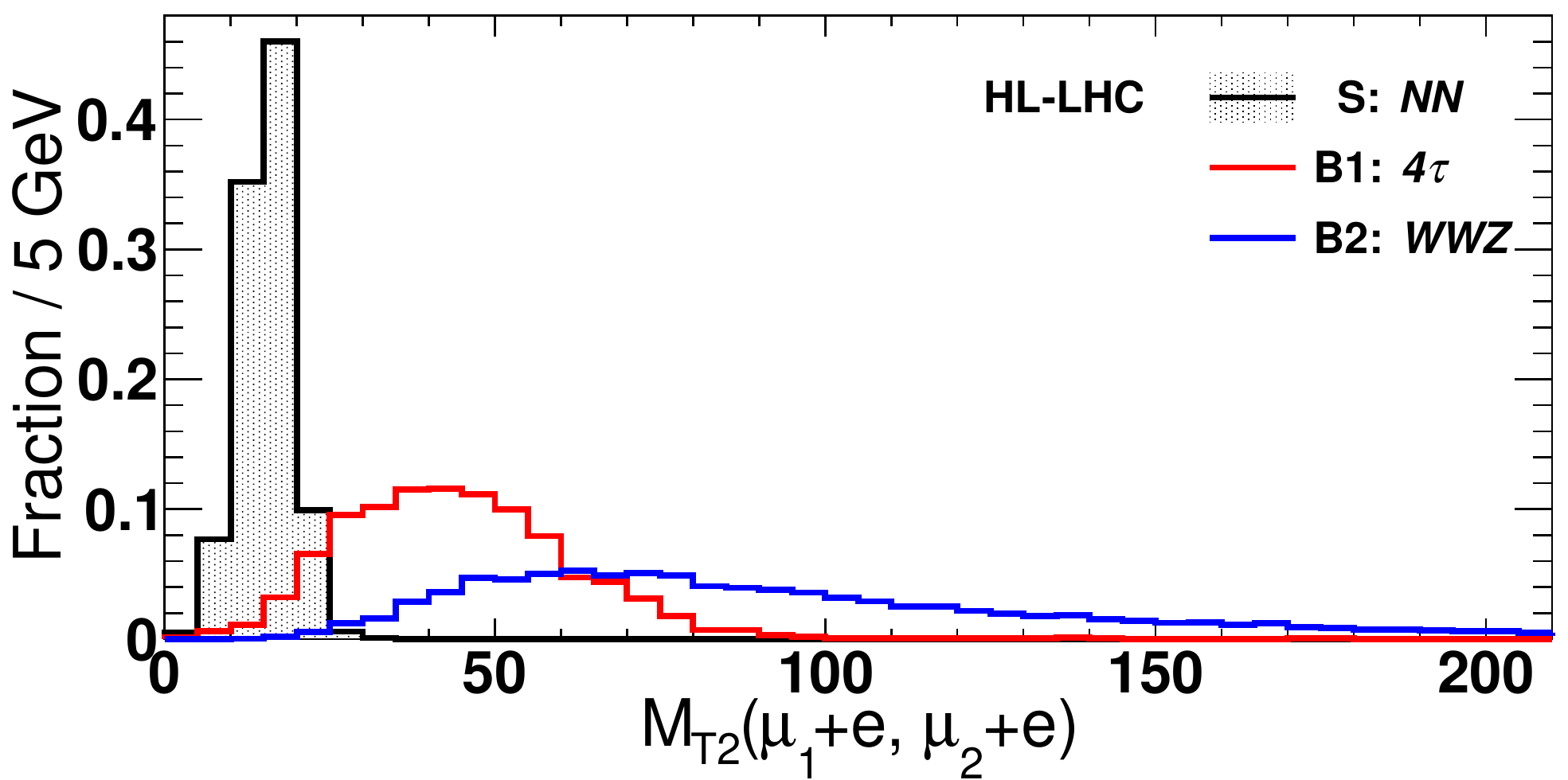}
\includegraphics[width=4.8cm,height=3cm]{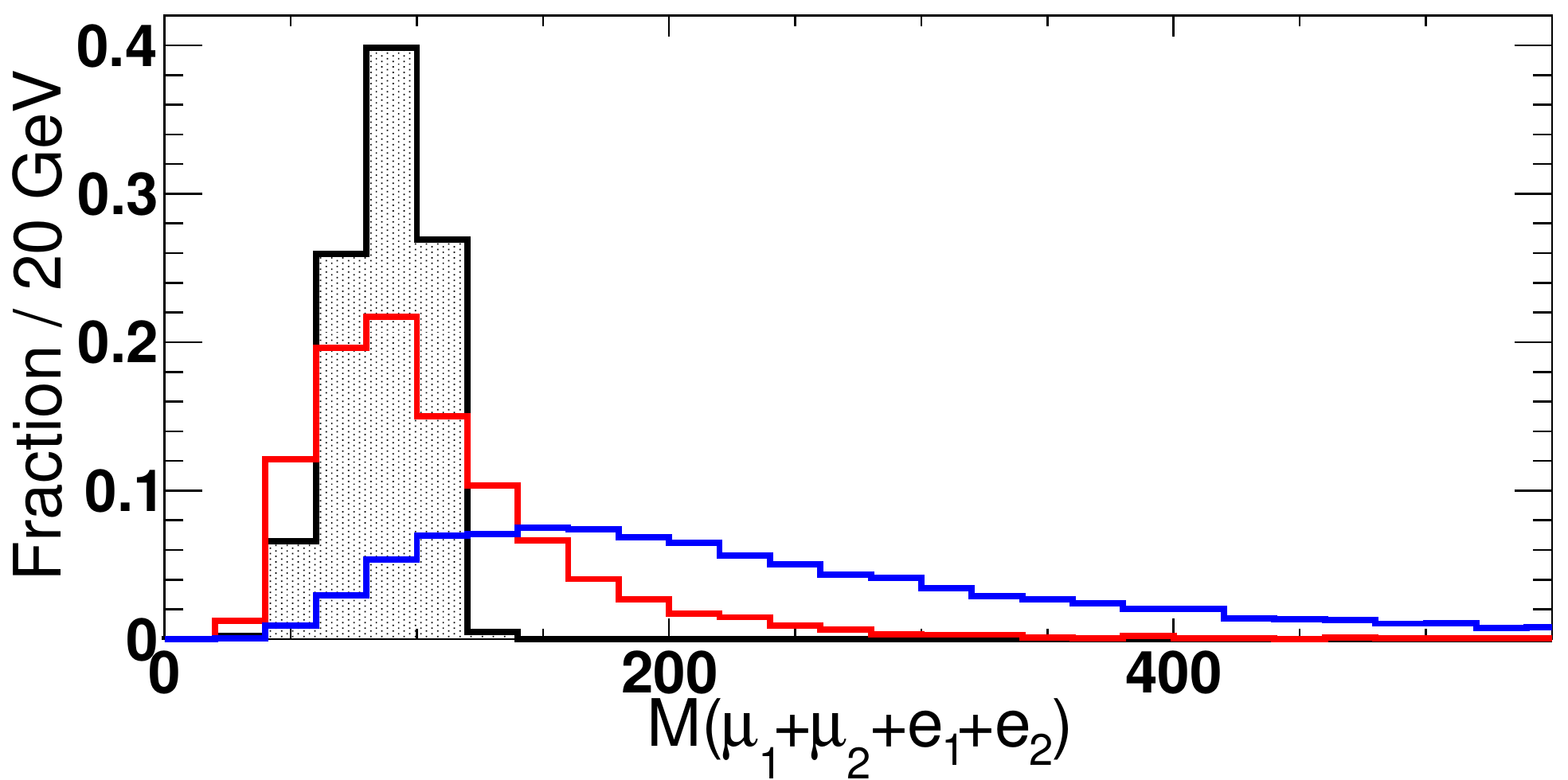}   
\includegraphics[width=4.8cm,height=3cm]{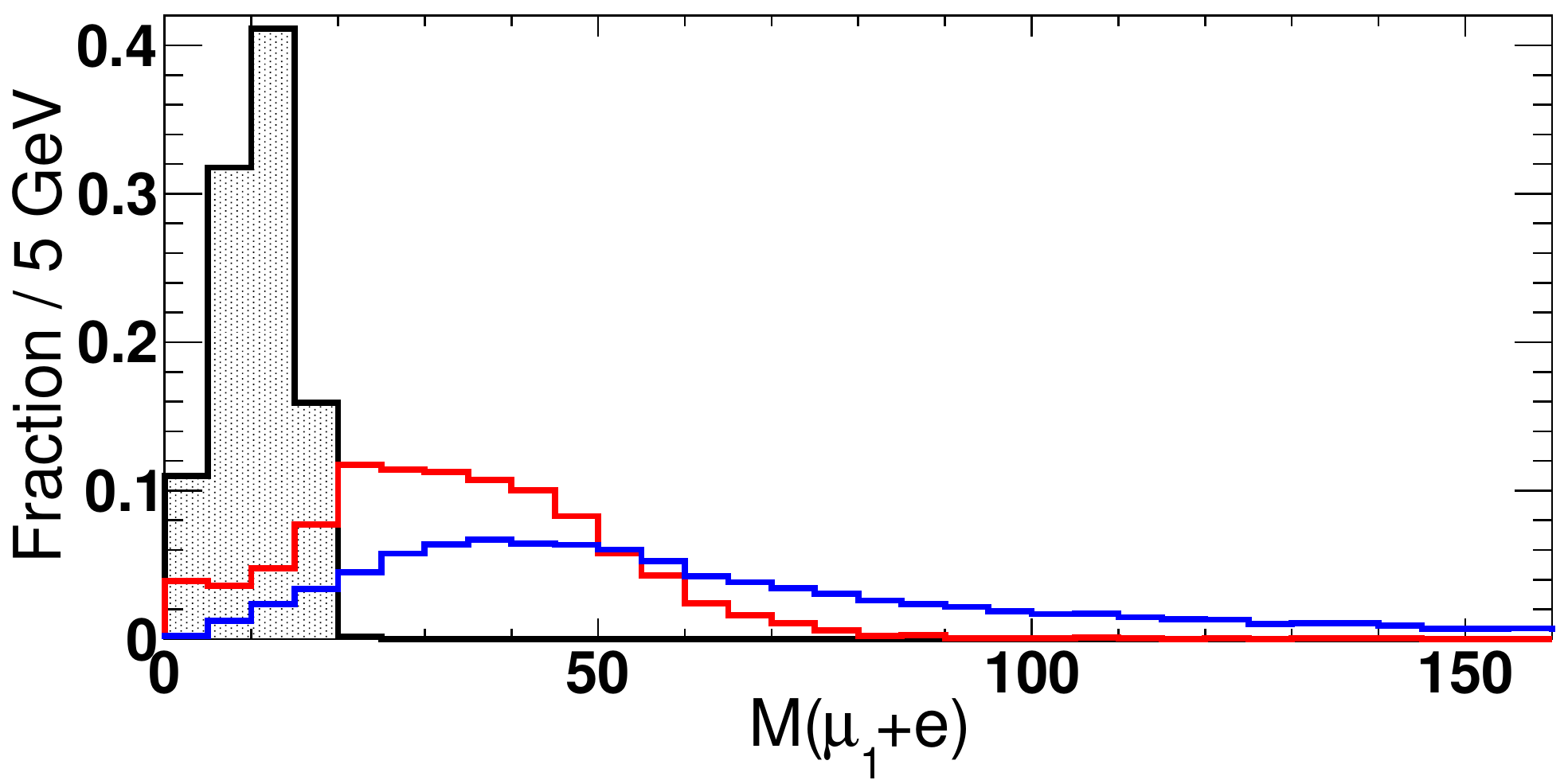}
\includegraphics[width=4.8cm,height=3cm]{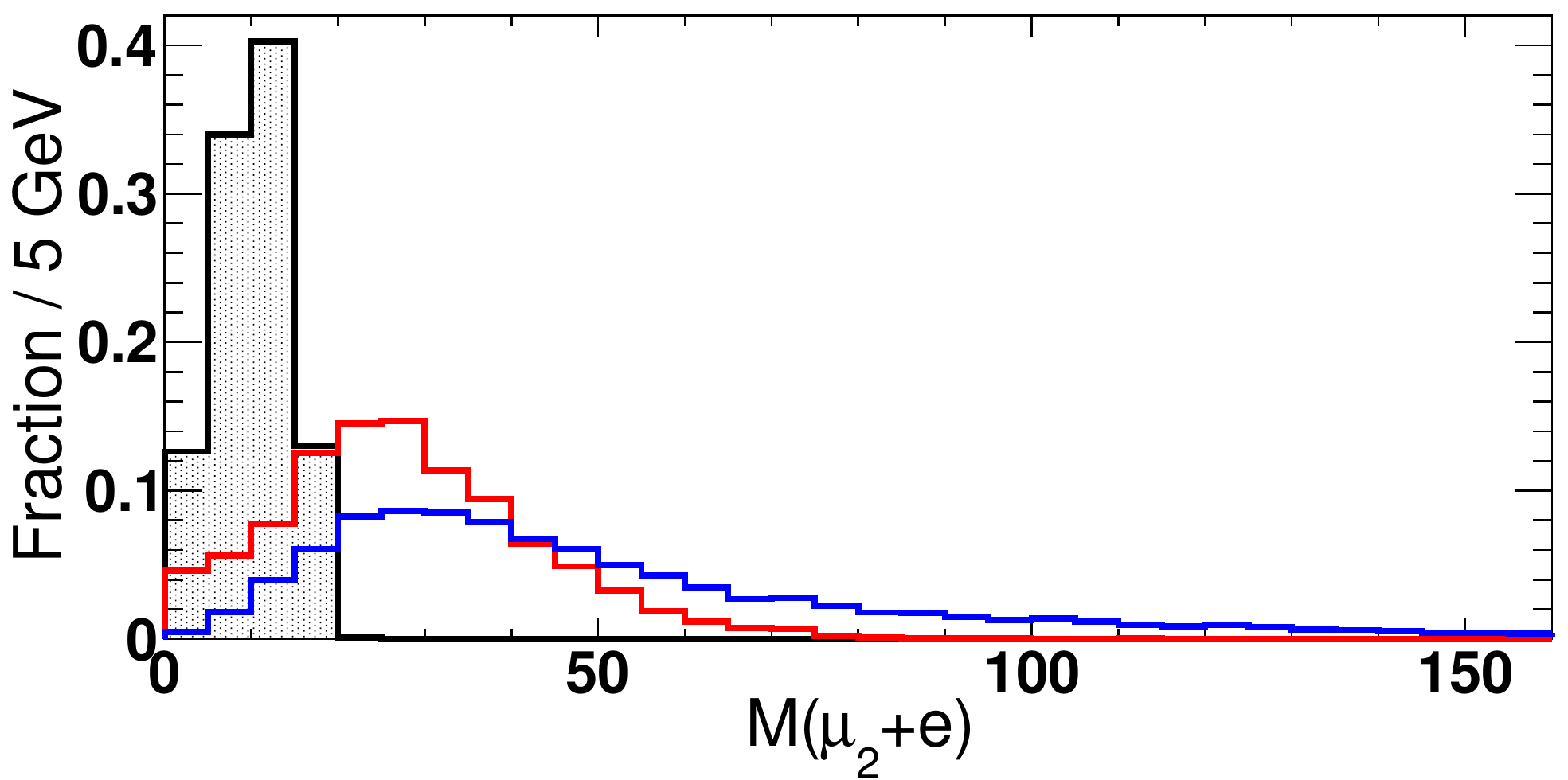}   
\includegraphics[width=4.8cm,height=3cm]{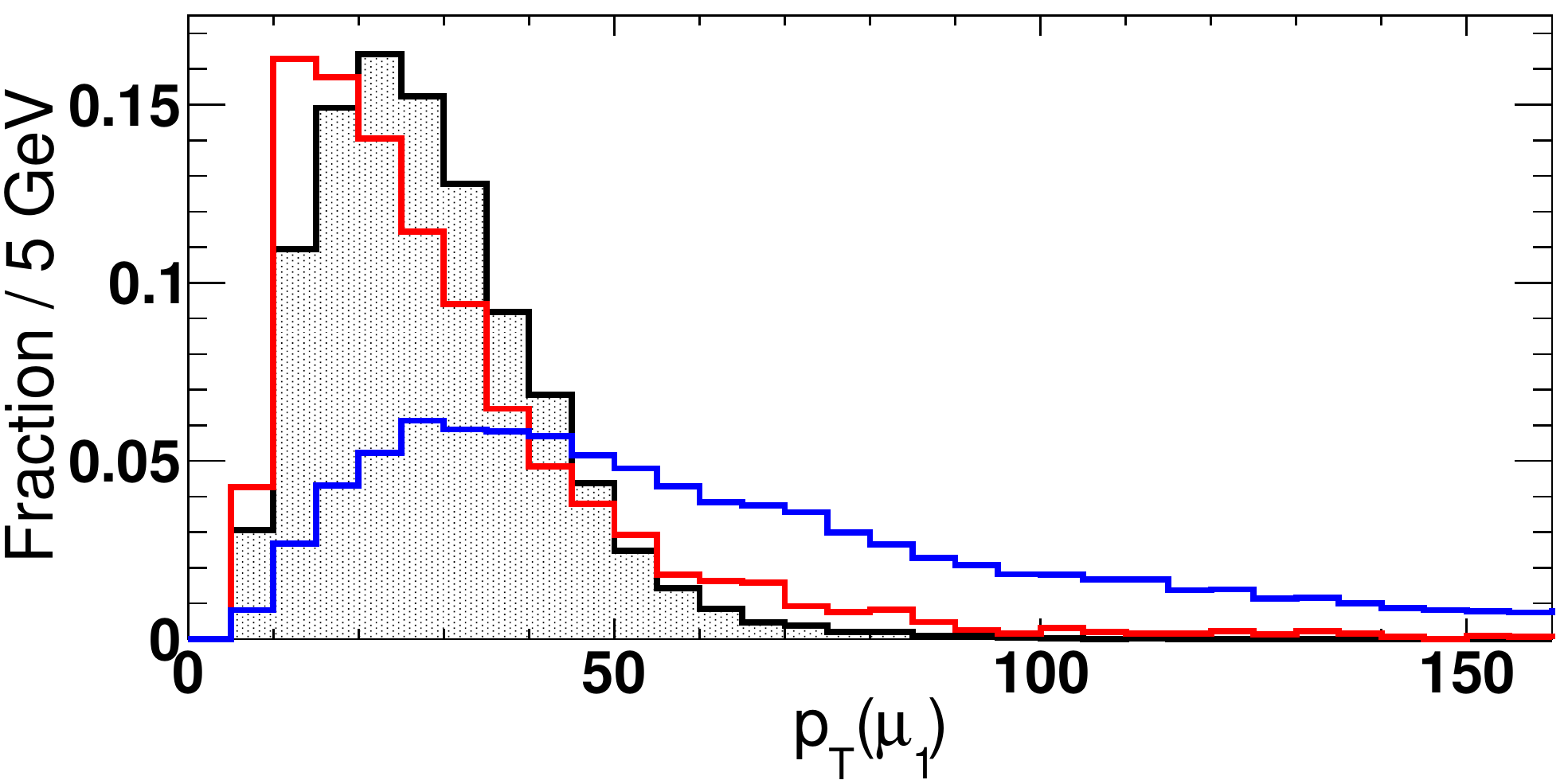}
\includegraphics[width=4.8cm,height=3cm]{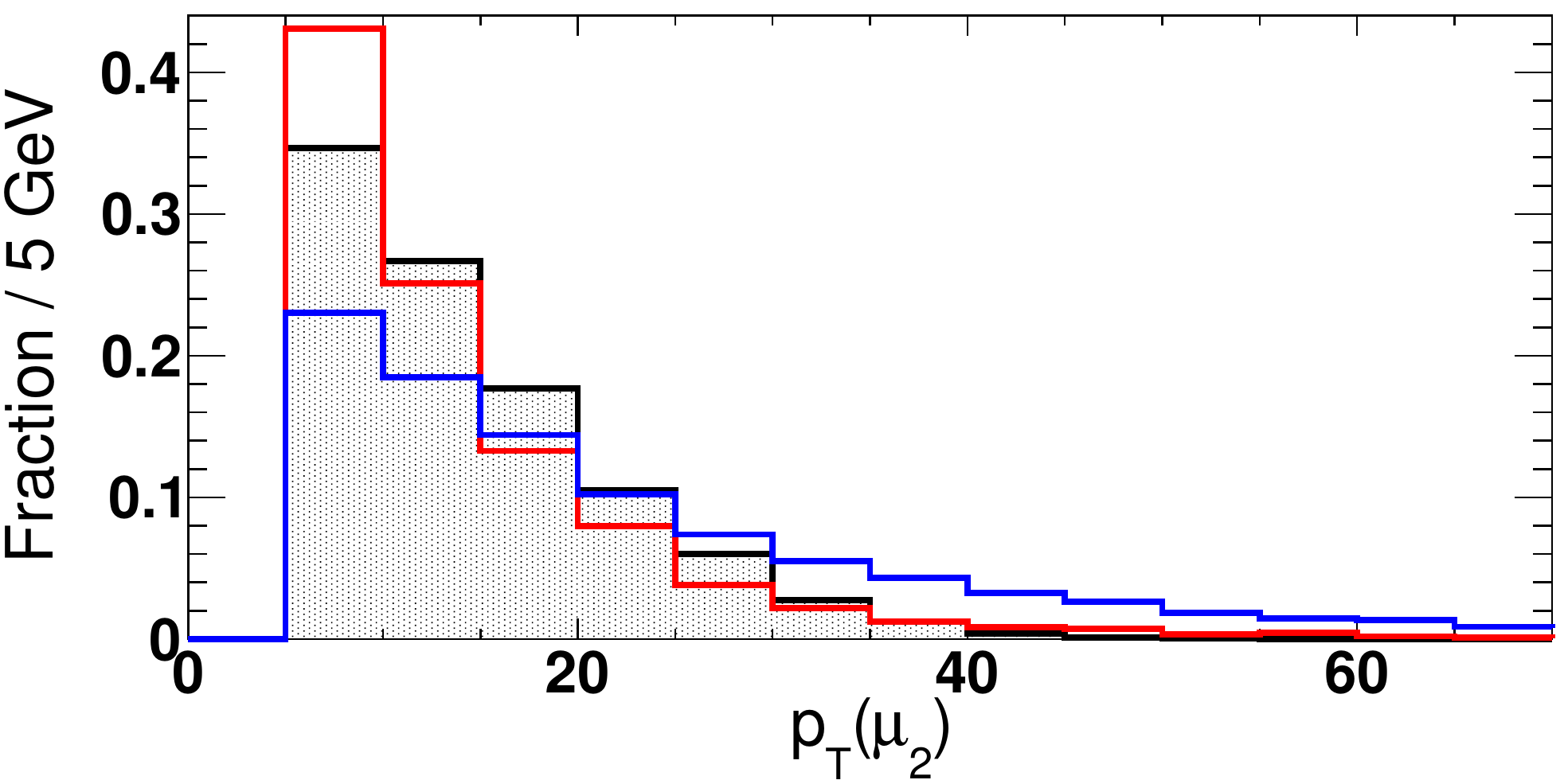}   
\includegraphics[width=4.8cm,height=3cm]{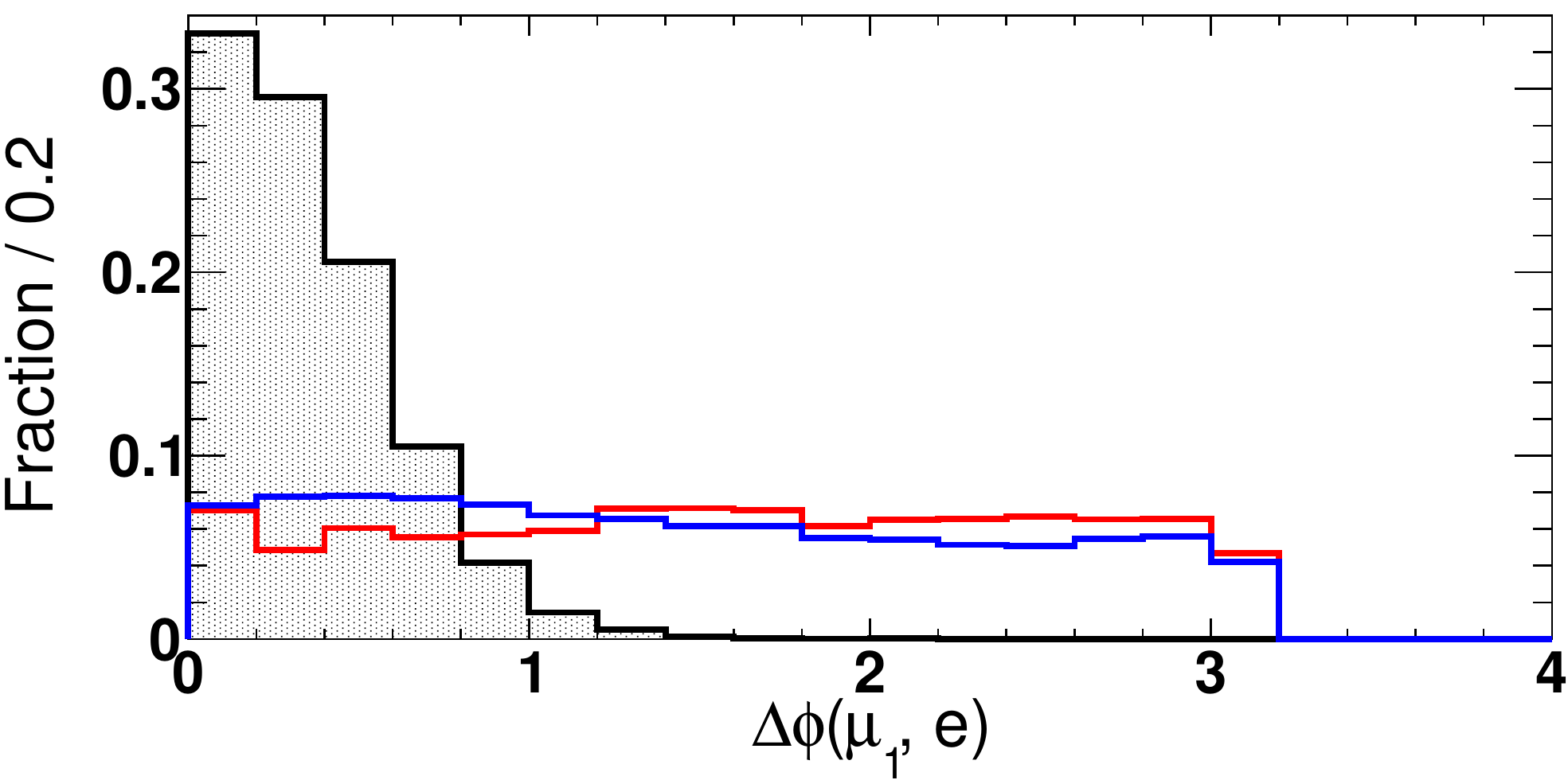}
\includegraphics[width=4.8cm,height=3cm]{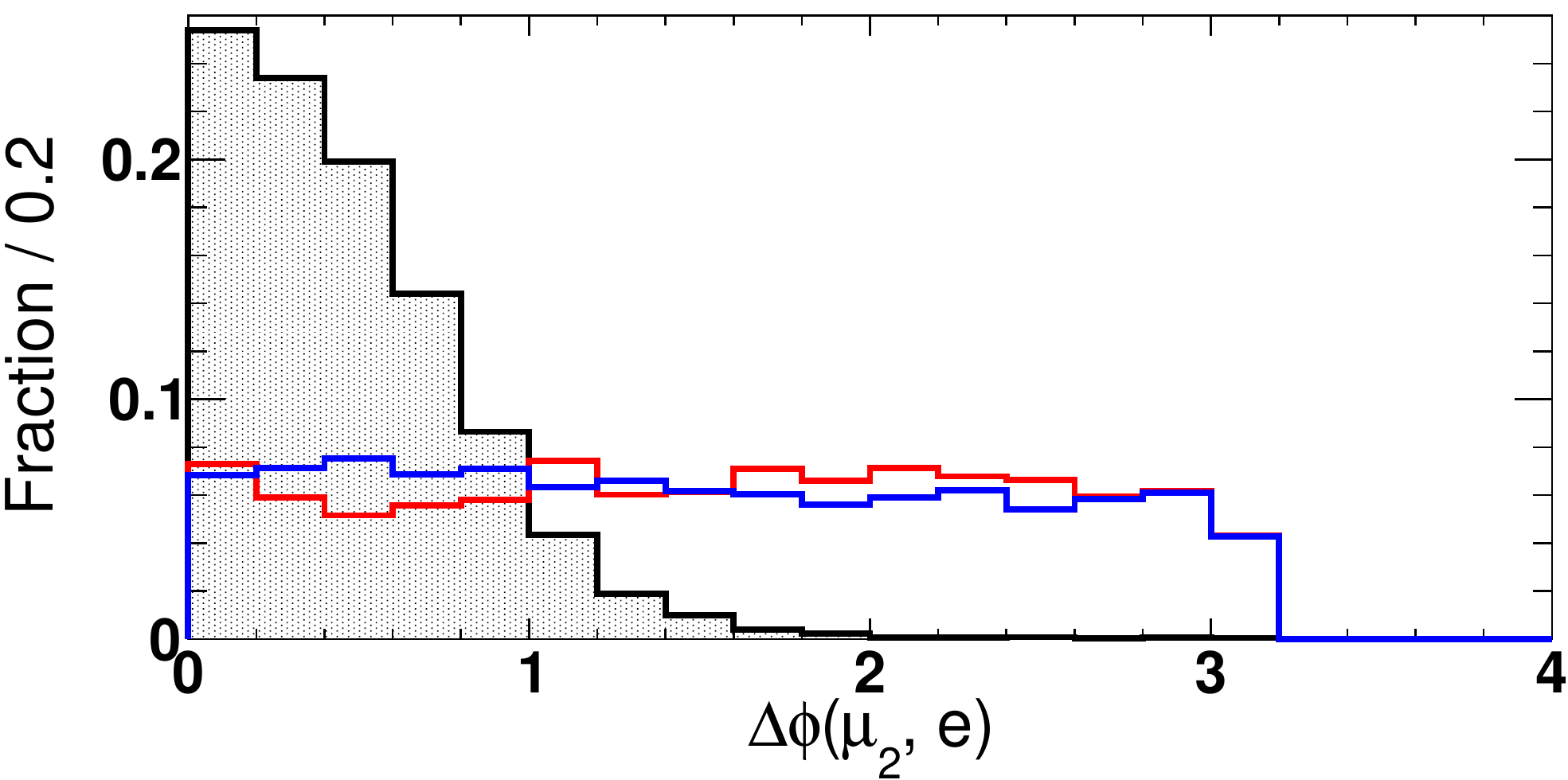}   
\includegraphics[width=4.8cm,height=3cm]{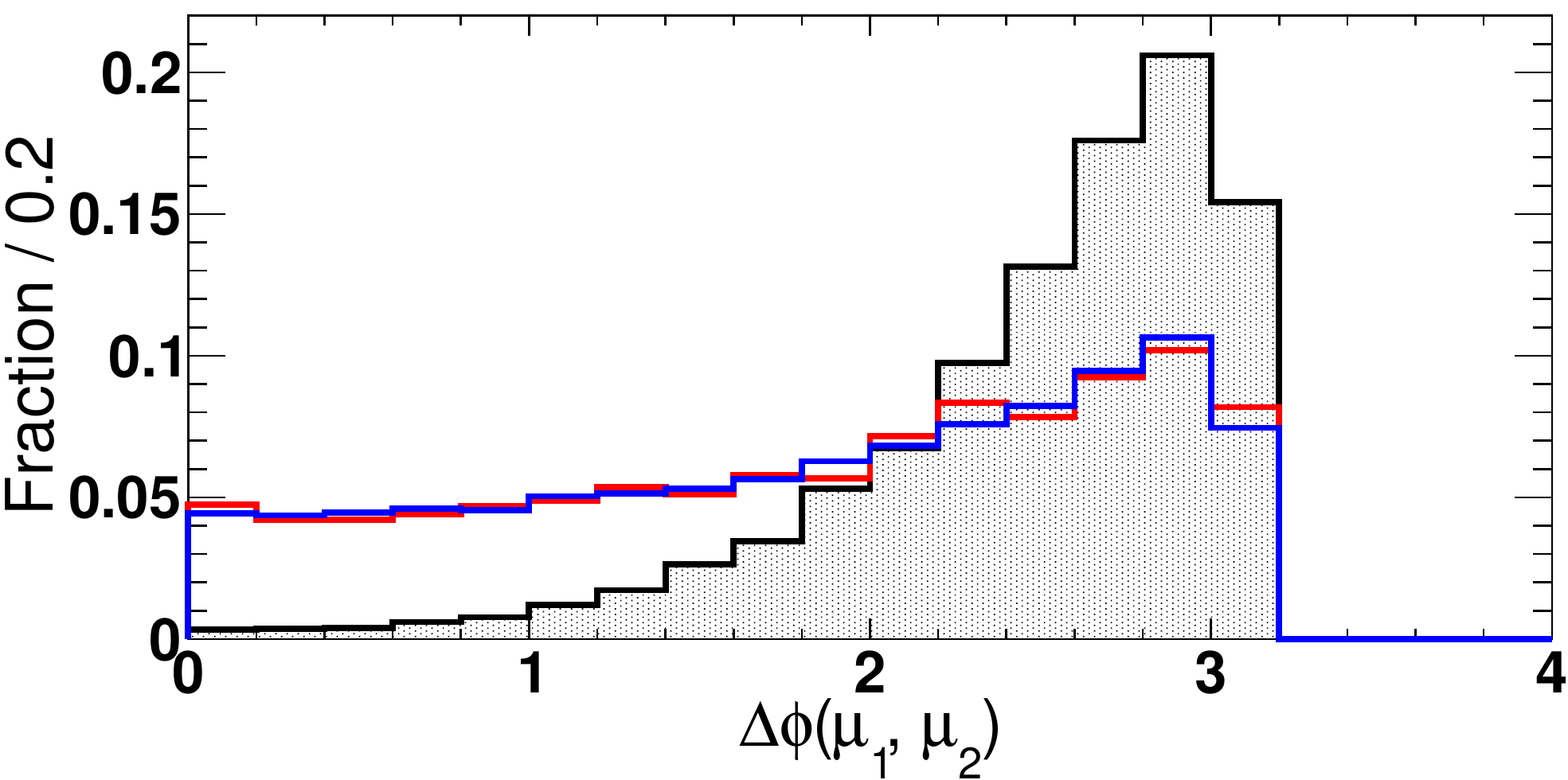}
\includegraphics[width=4.8cm,height=3cm]{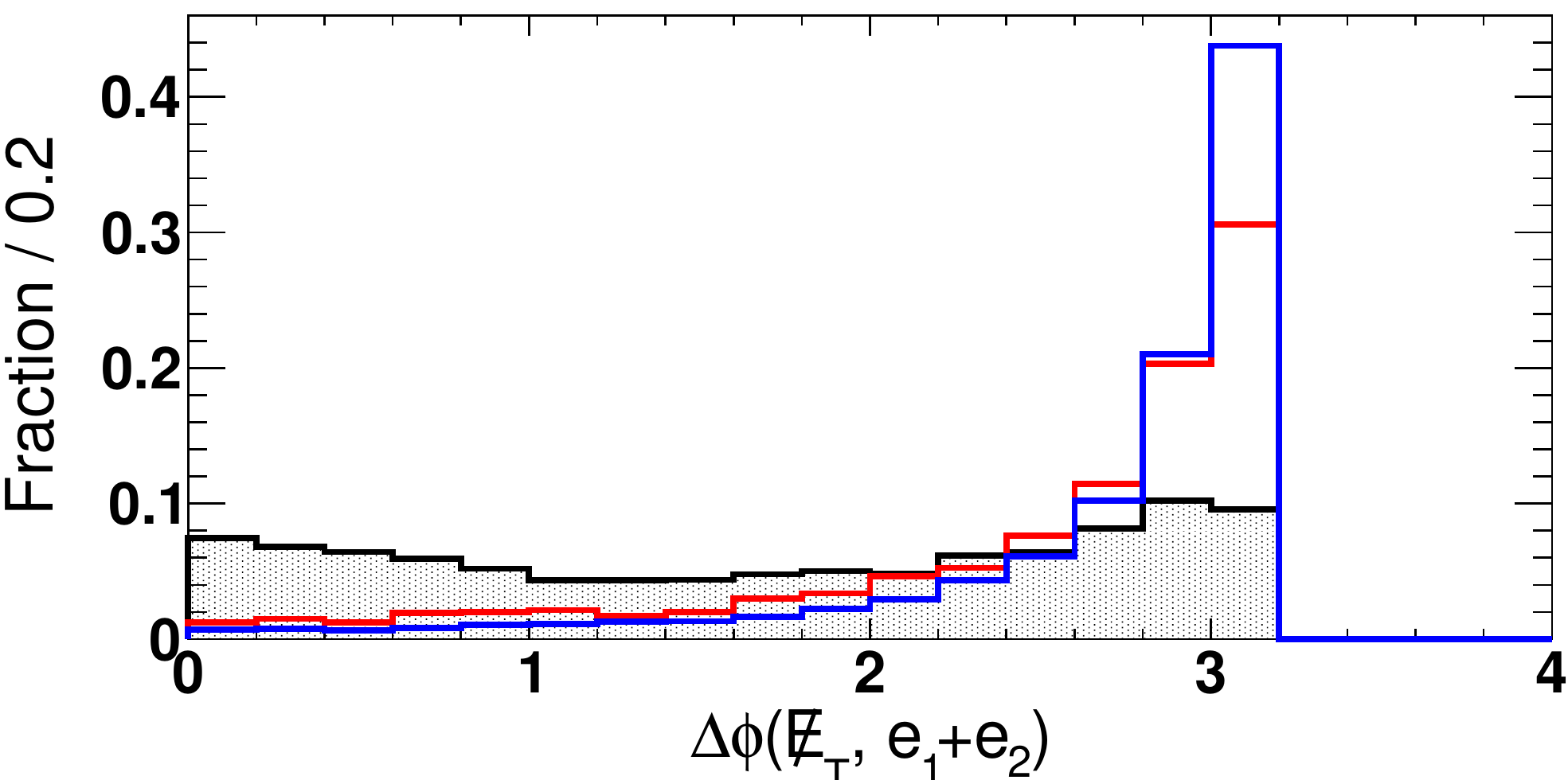}   
\includegraphics[width=4.8cm,height=3cm]{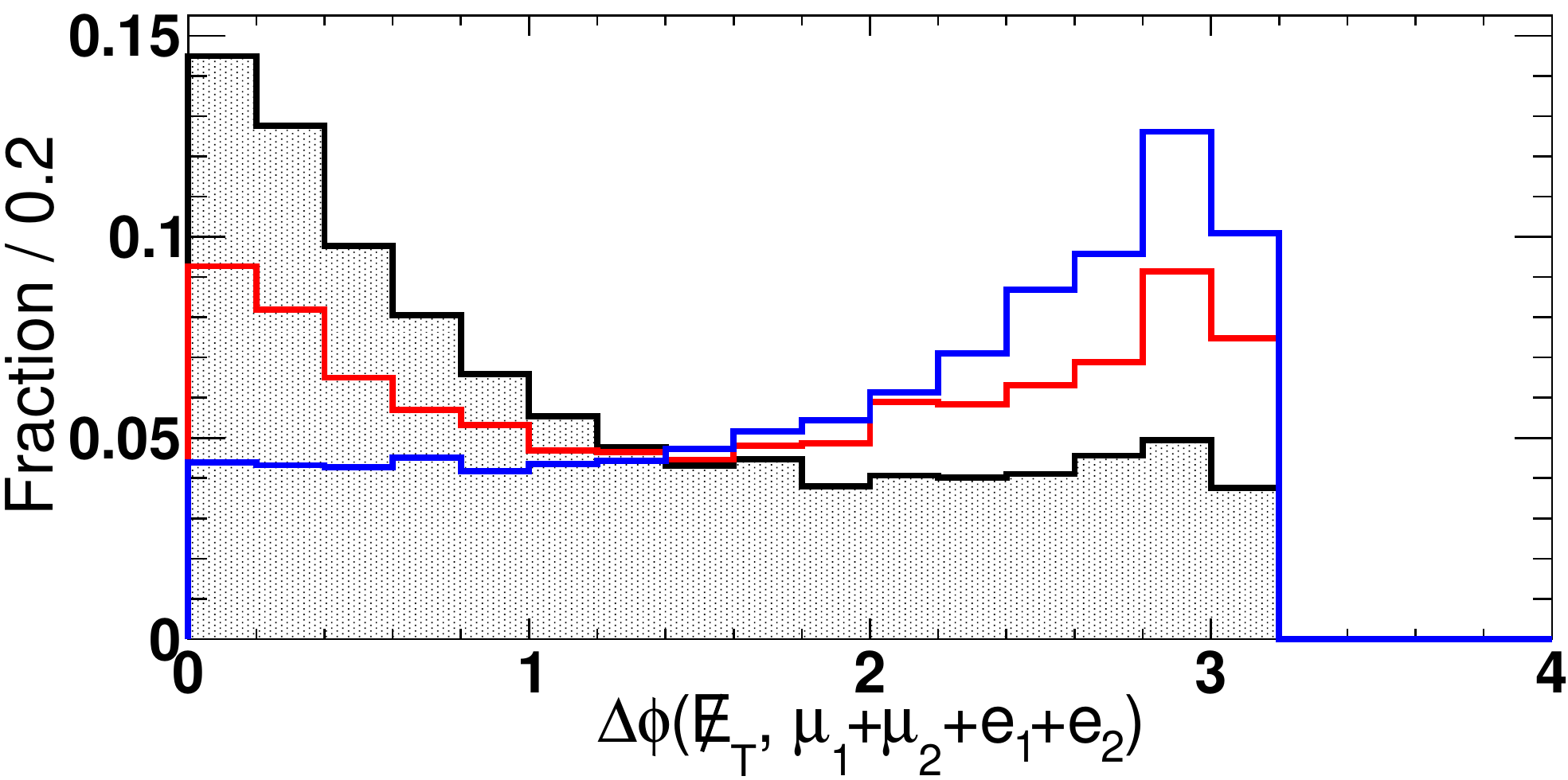} 
\includegraphics[width=4.8cm,height=3cm]{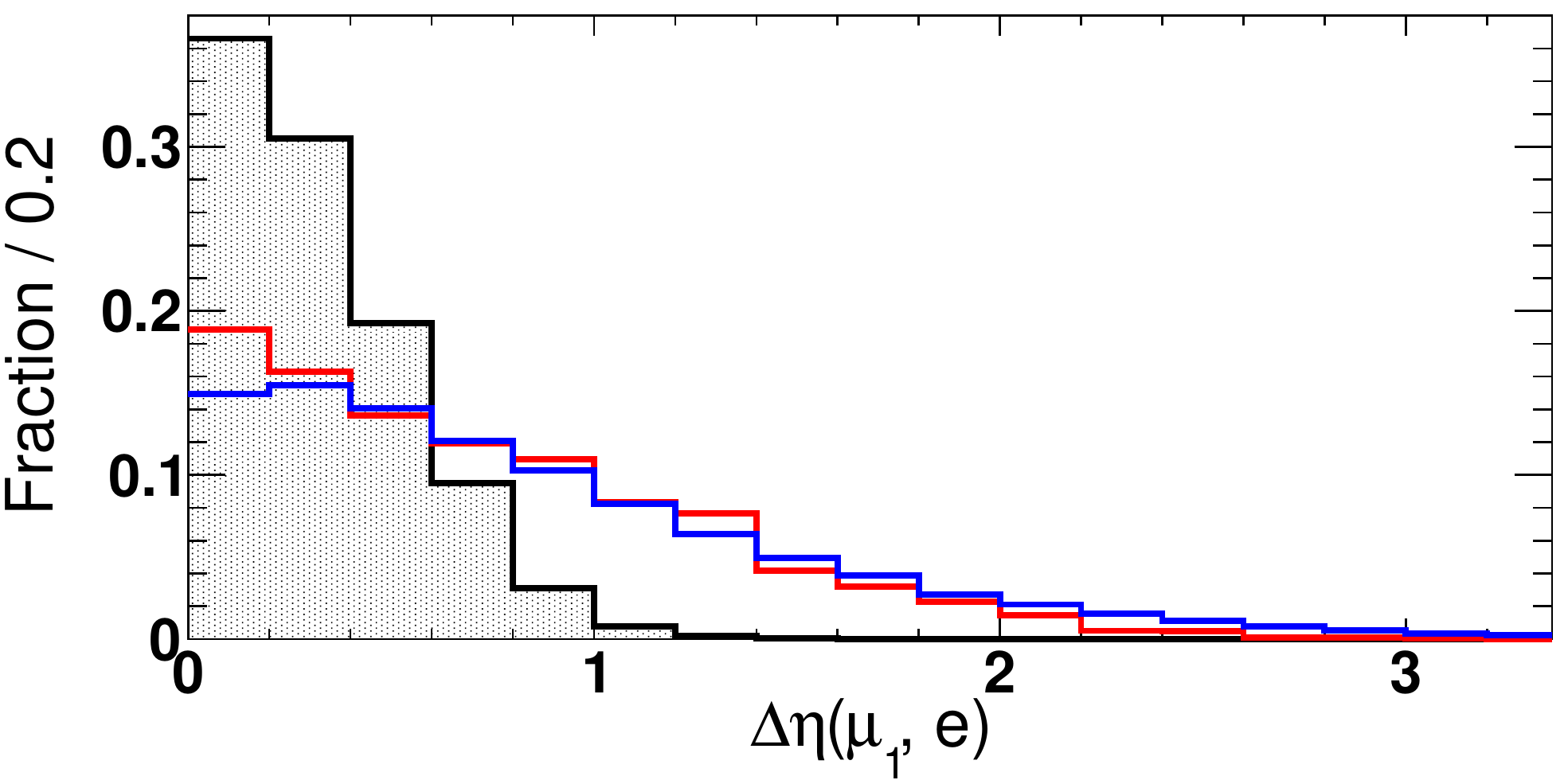}
\caption{Distributions of some input observables for the signal event sample (black, filled), and those for the SM background samples including $4\tau$ (red) and $WWZ$ (blue). This plot illustrates a $h_1$ decay signal by gluon fusion at the HL-LHC, with $m_N$=20 GeV.}
\label{fig:ObsmN20}
\end{figure}

\section{Distributions of BDT responses}
\label{appendix:BDTdistribution}

In Fig.~\ref{fig:BDT14TeV}, we show the distributions of BDT responses of the signal and the SM background processes at the HL-LHC with $\sqrt{s} = 14$ TeV for different $m_N$ assumptions. In these plots, the signals are from $h_1$ decay only. The distributions at the FCC-hh/SppC with $\sqrt{s} = 100$ TeV are presented in Fig.~\ref{fig:BDT100TeV}.

\newpage
\begin{figure}[h]
\centering
\includegraphics[width=4.8cm,height=3cm]{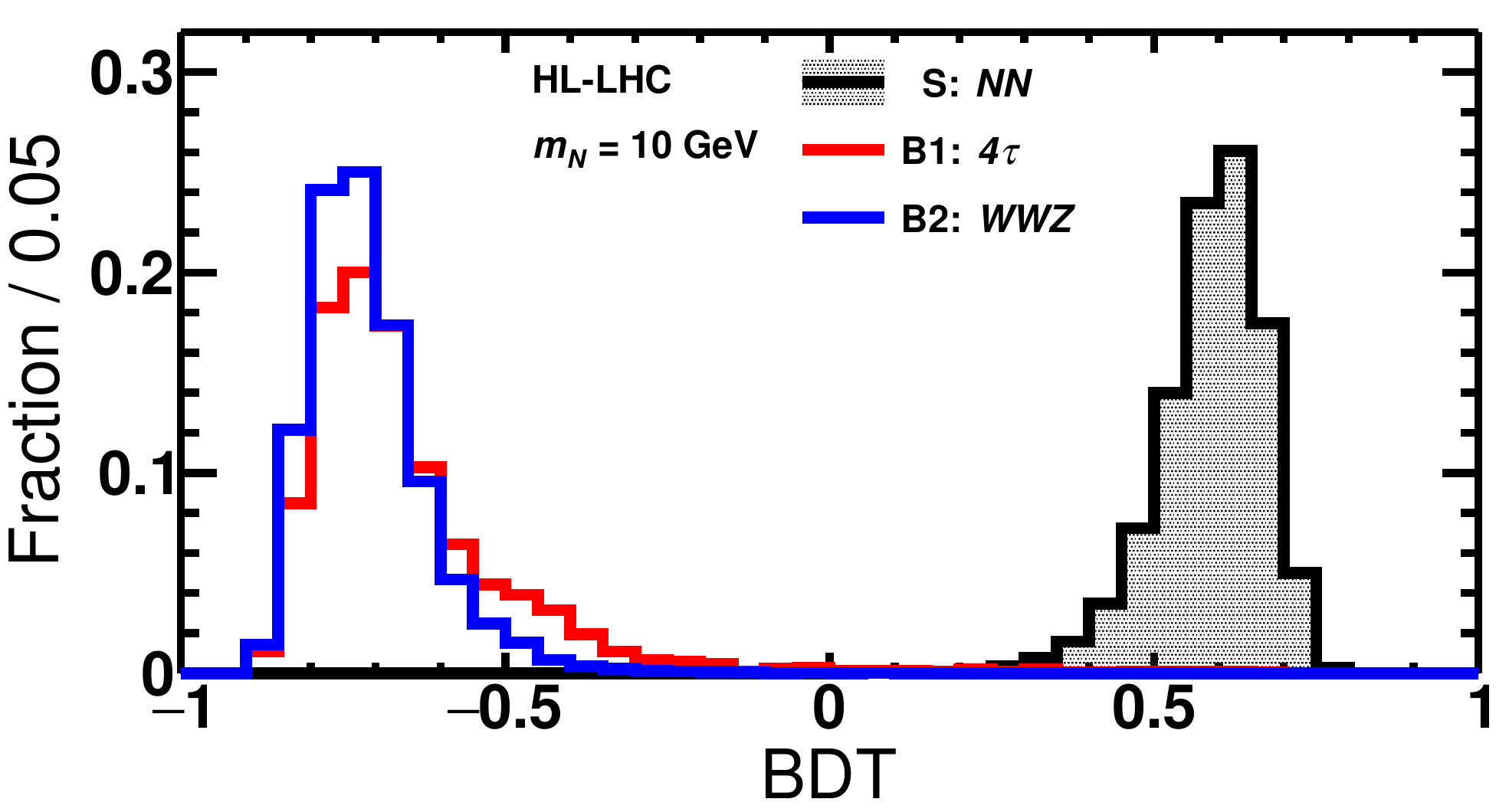}
\includegraphics[width=4.8cm,height=3cm]{fig_BDT_14TeV_mN20.pdf}   
\includegraphics[width=4.8cm,height=3cm]{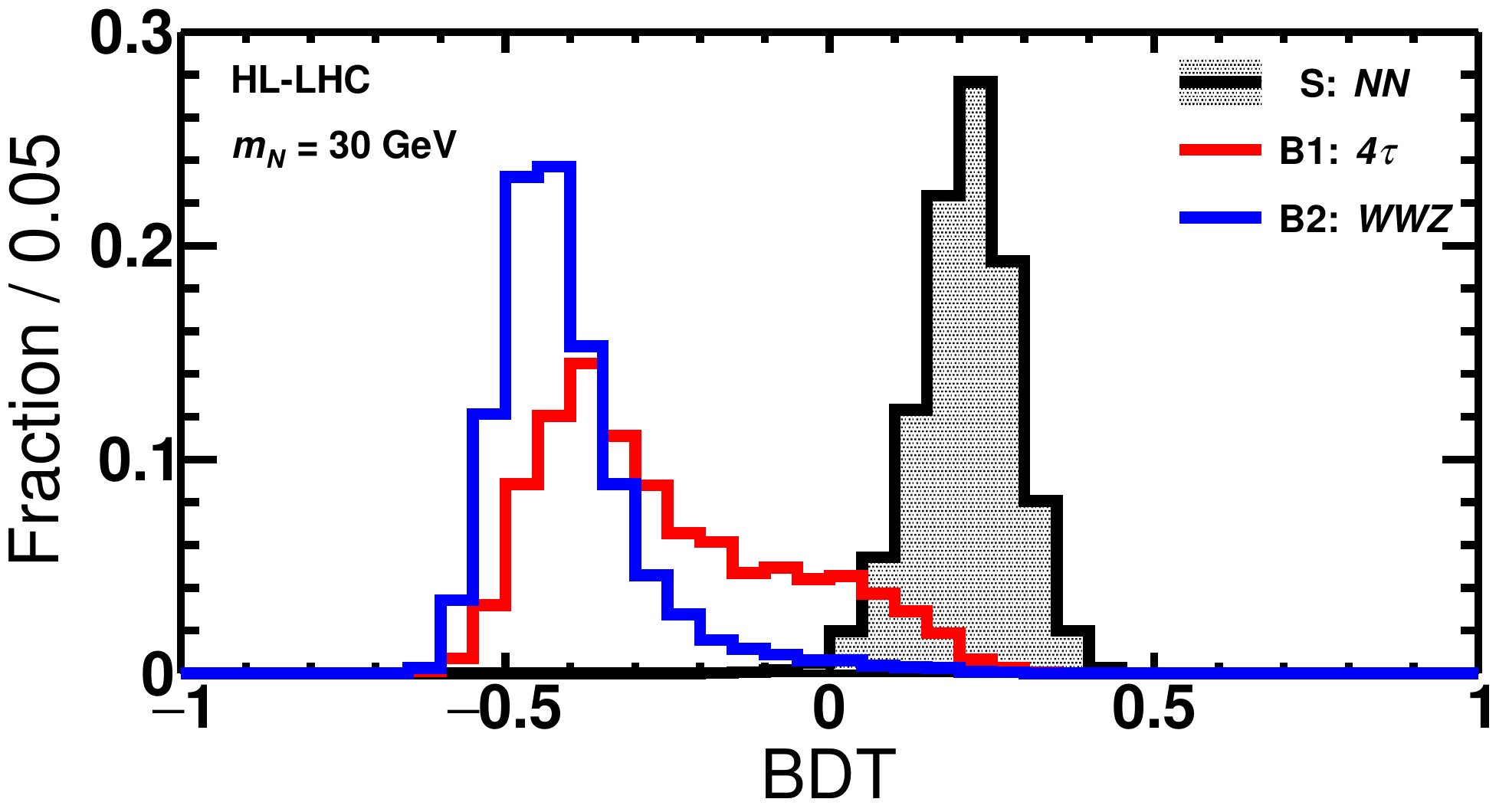}
\includegraphics[width=4.8cm,height=3cm]{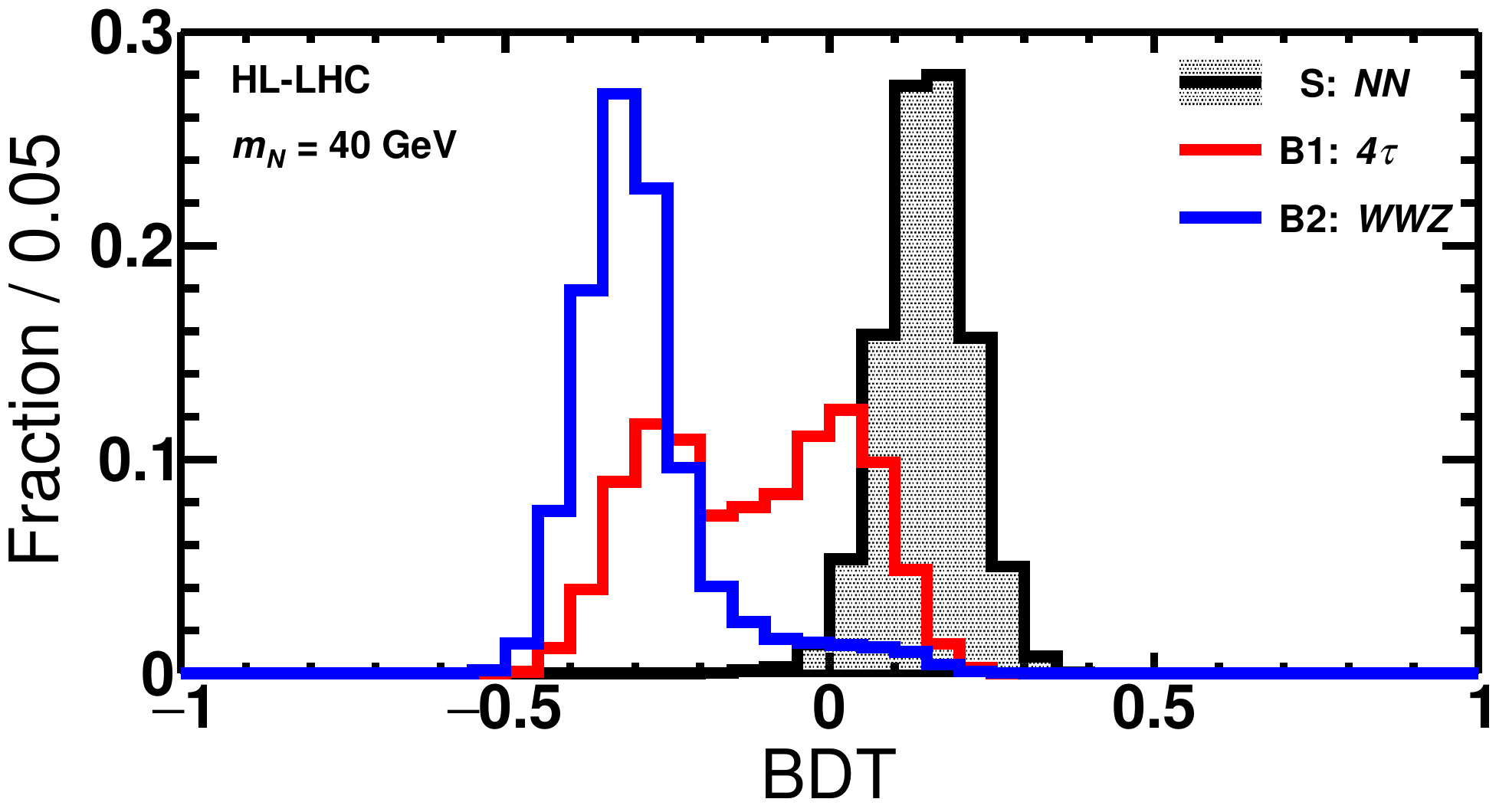}   
\includegraphics[width=4.8cm,height=3cm]{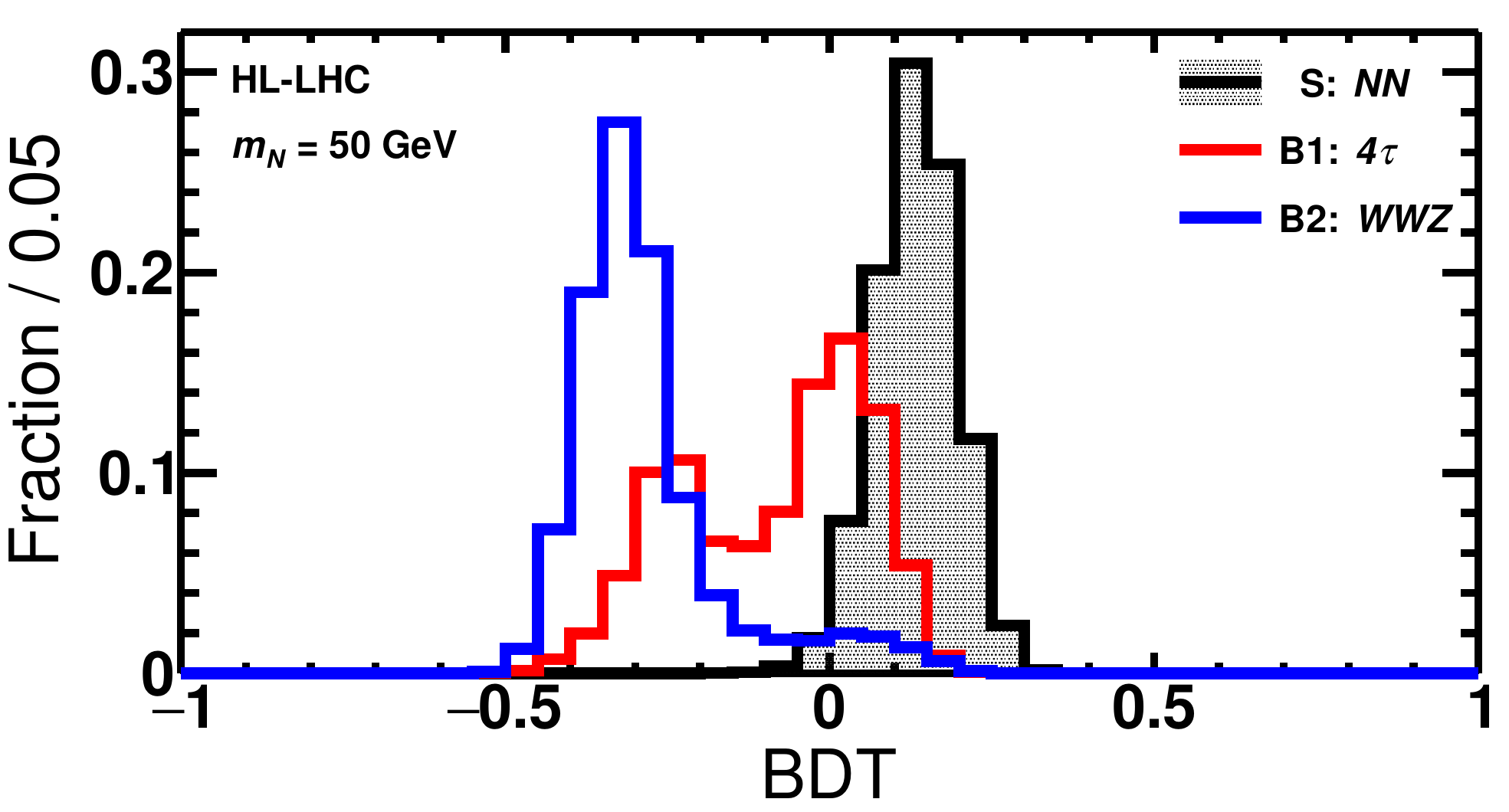}
\includegraphics[width=4.8cm,height=3cm]{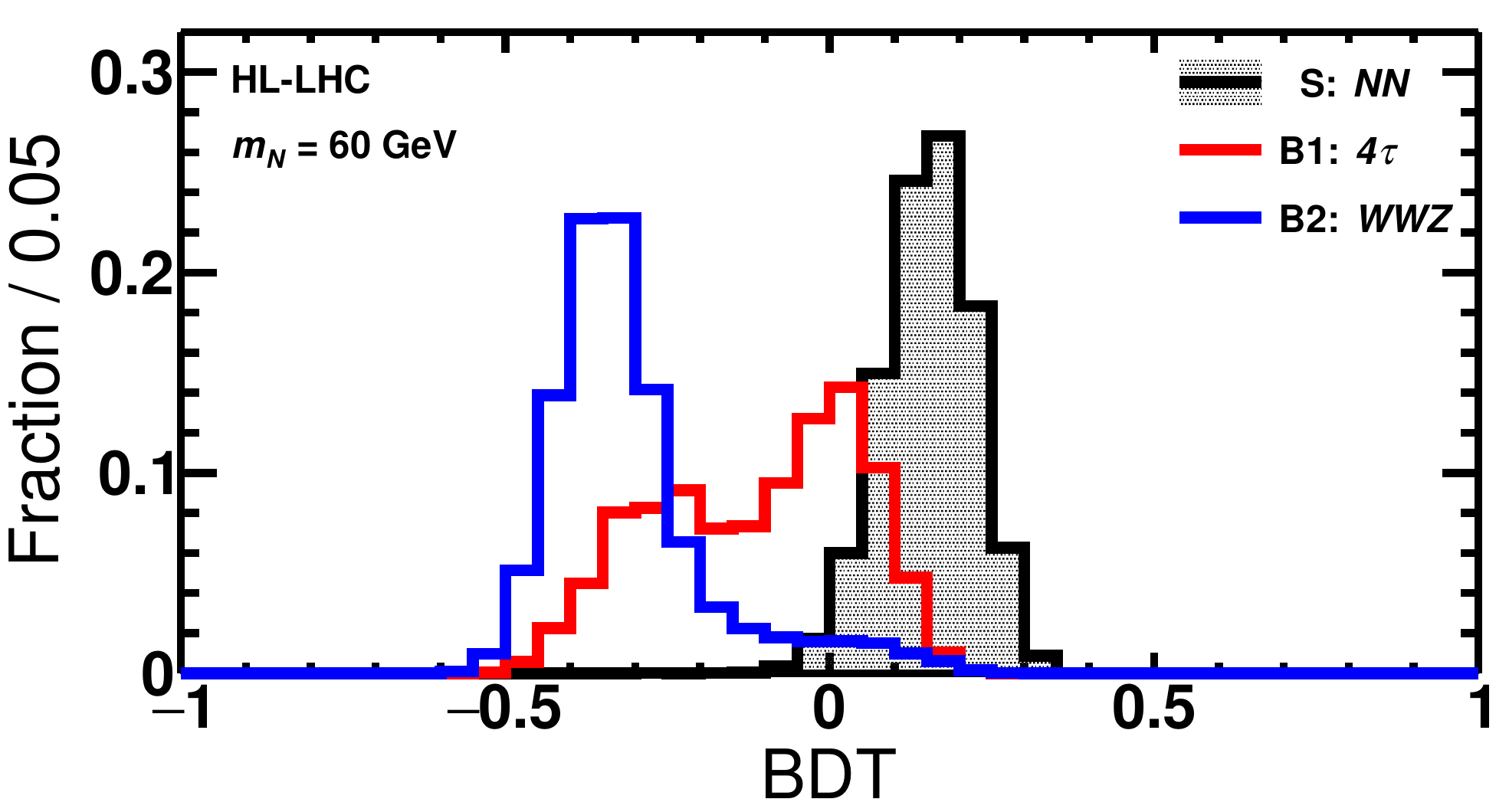}   
\caption{Distributions of BDT responses for the signal event sample (black, filled), and those for the SM background samples including $4\tau$ (red) and $WWZ$ (blue).
This figure illustrates an $h_1$ decay signal by gluon fusion at the HL-LHC, with different $m_N$ assumptions.}
\label{fig:BDT14TeV}
\end{figure}

\begin{figure}[h]
\centering
\includegraphics[width=4.8cm,height=3cm]{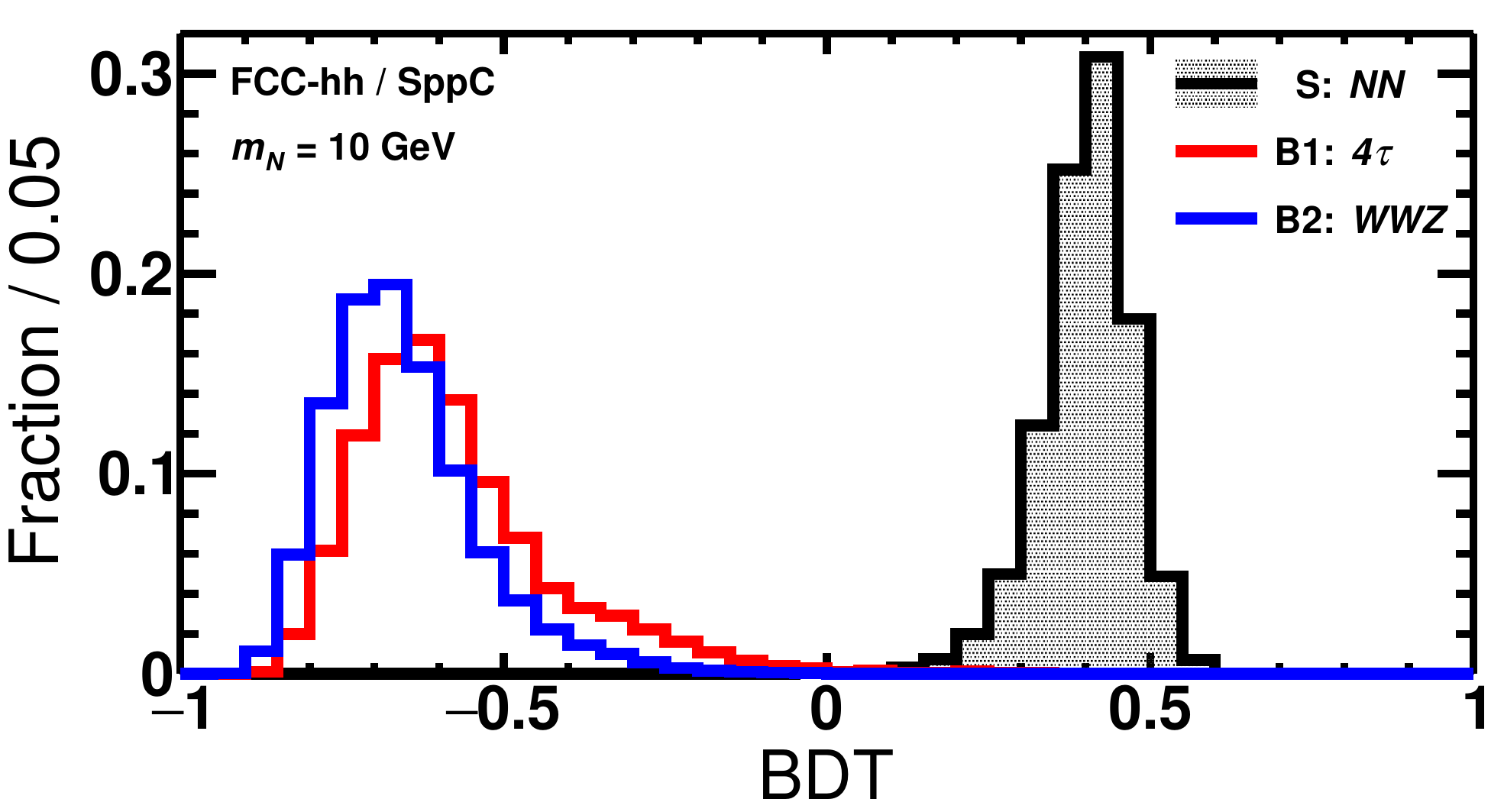}
\includegraphics[width=4.8cm,height=3cm]{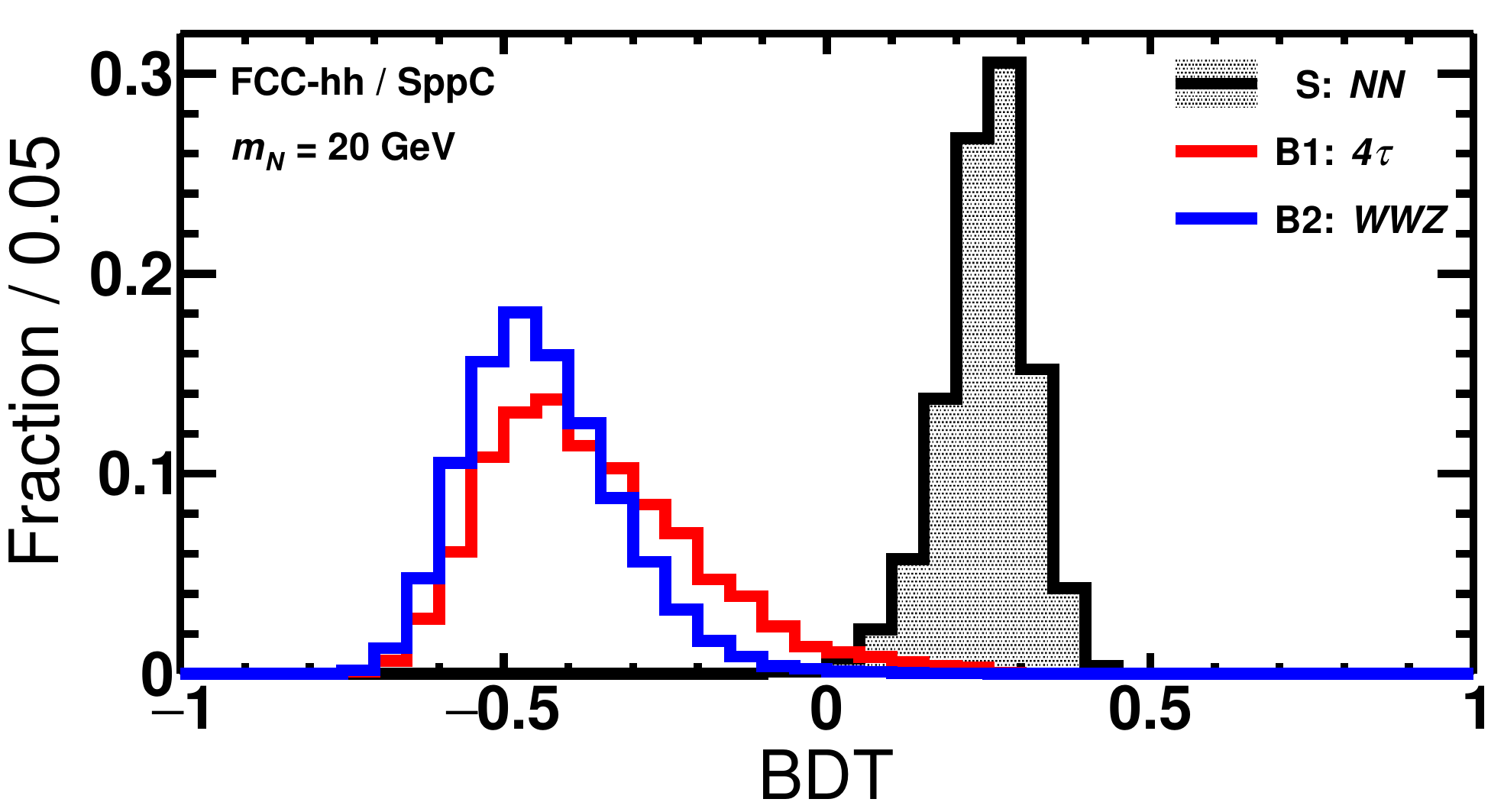}   
\includegraphics[width=4.8cm,height=3cm]{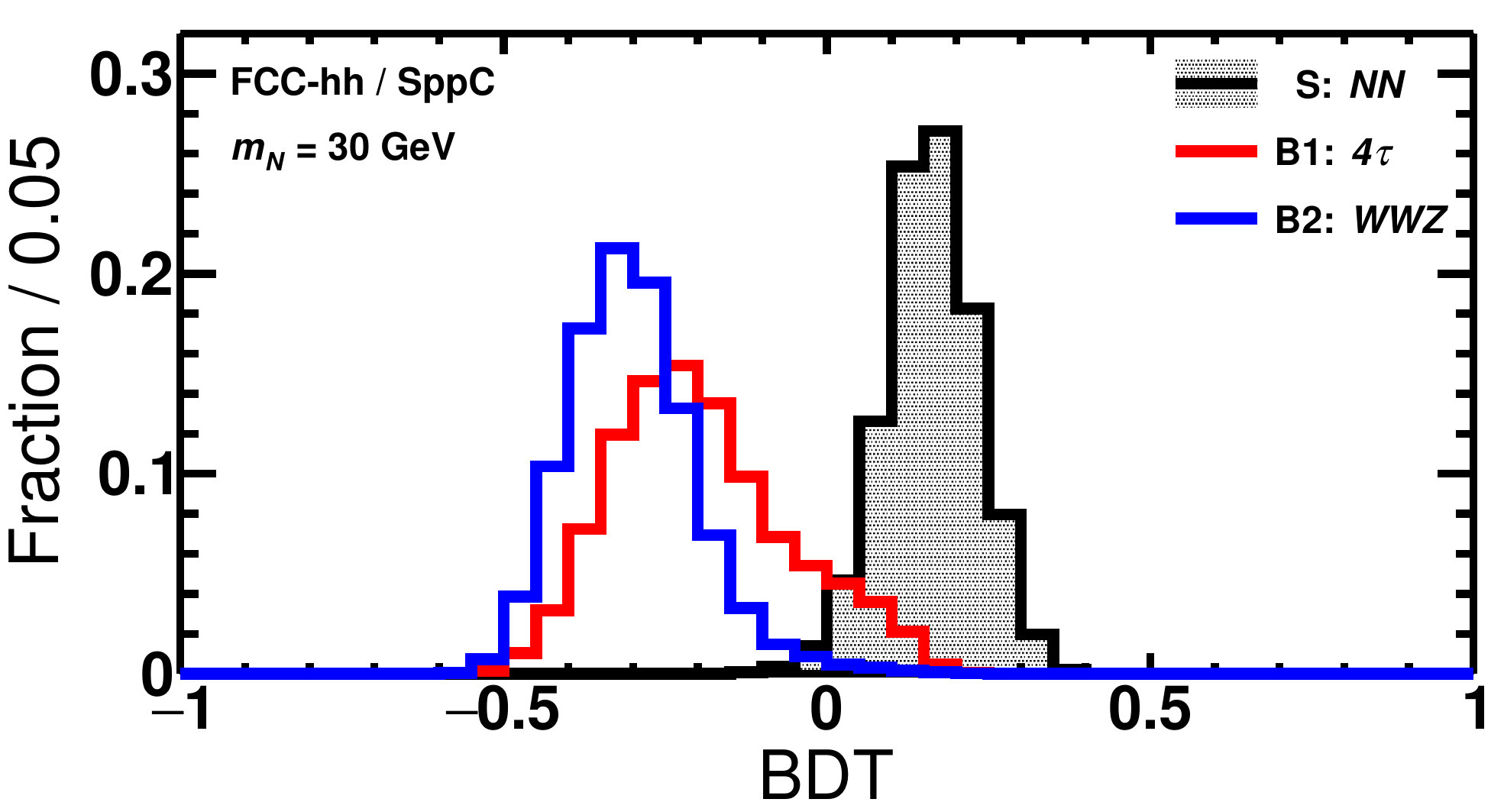}
\includegraphics[width=4.8cm,height=3cm]{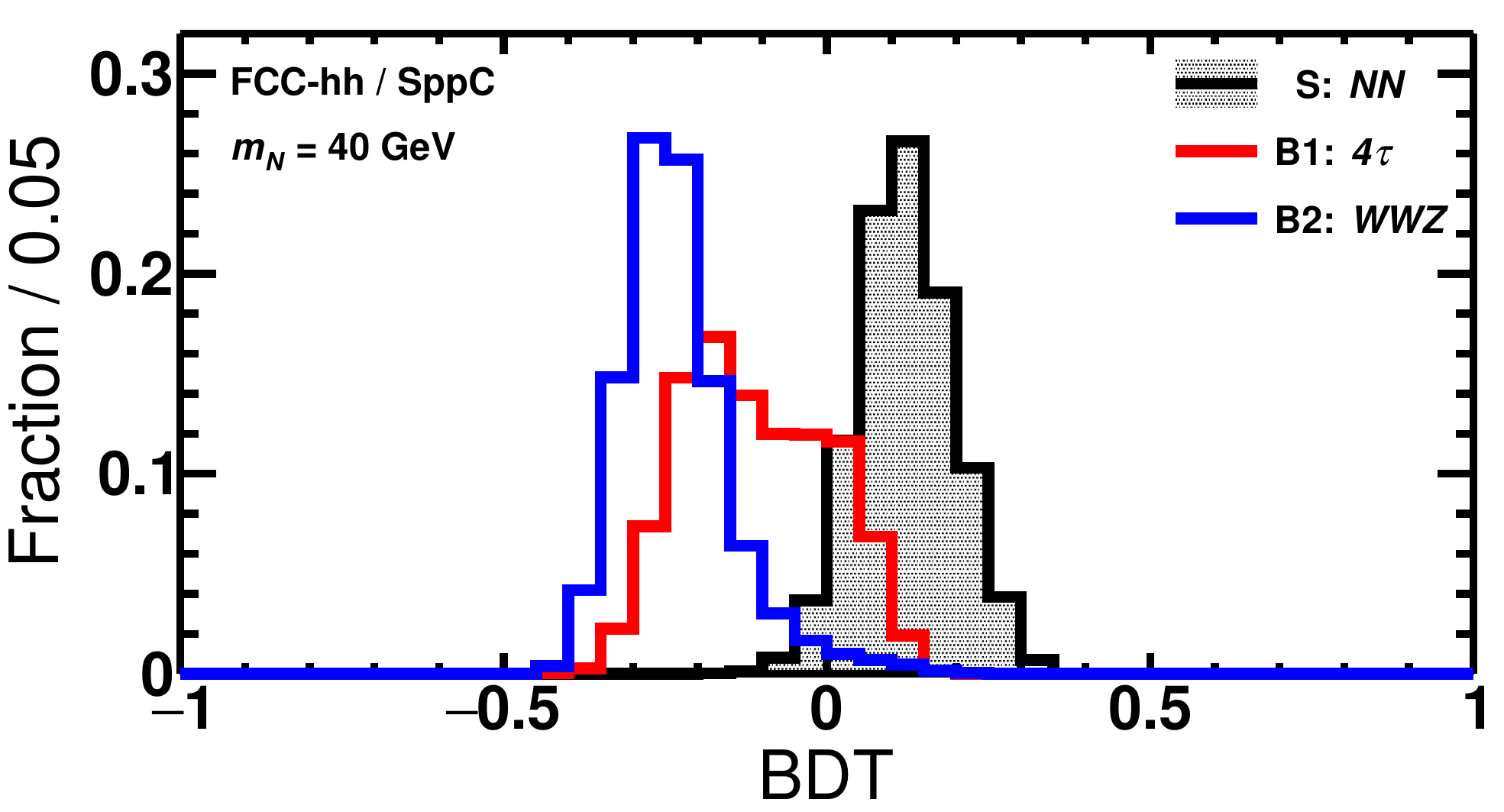}   
\includegraphics[width=4.8cm,height=3cm]{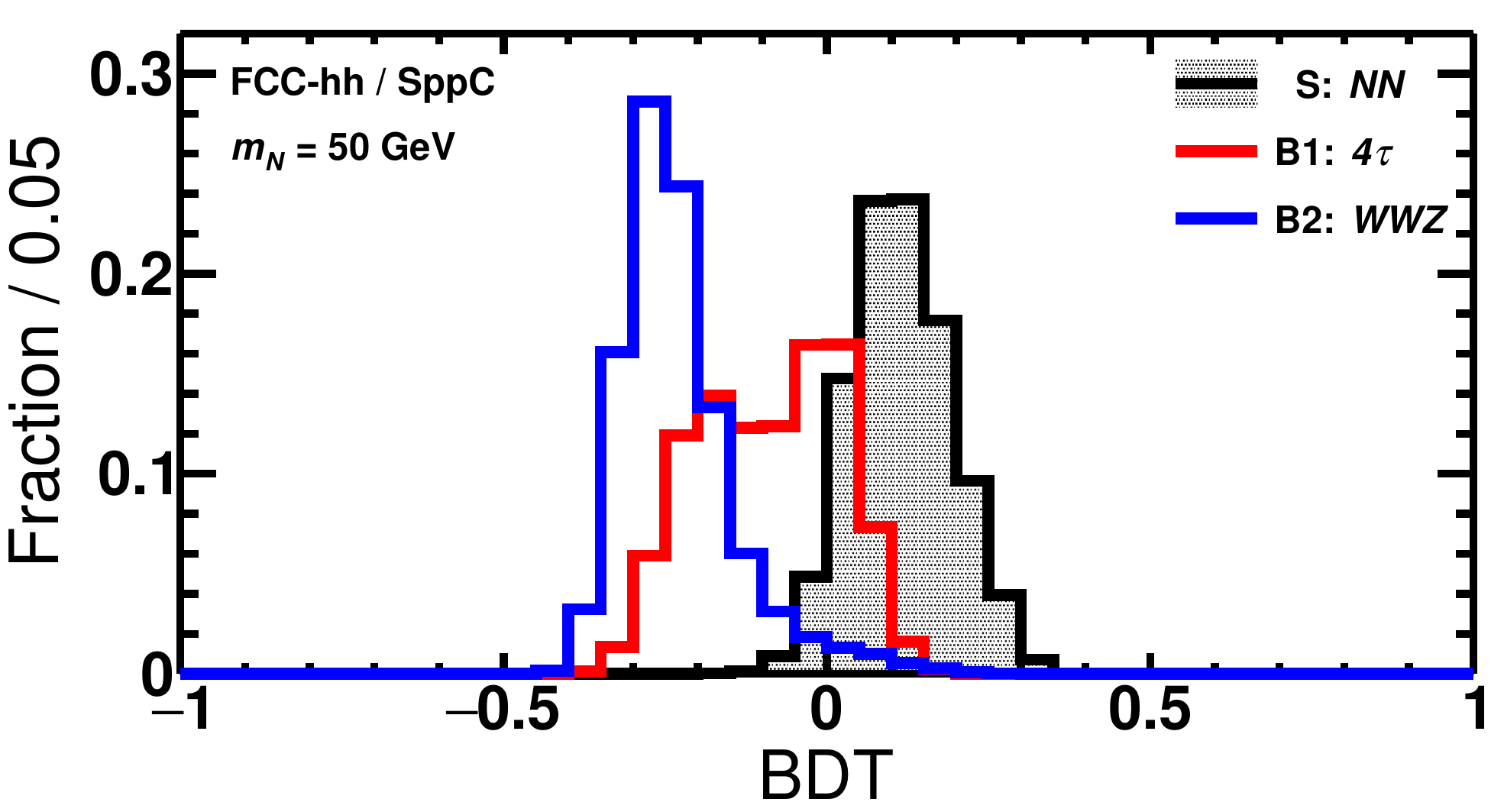}
\includegraphics[width=4.8cm,height=3cm]{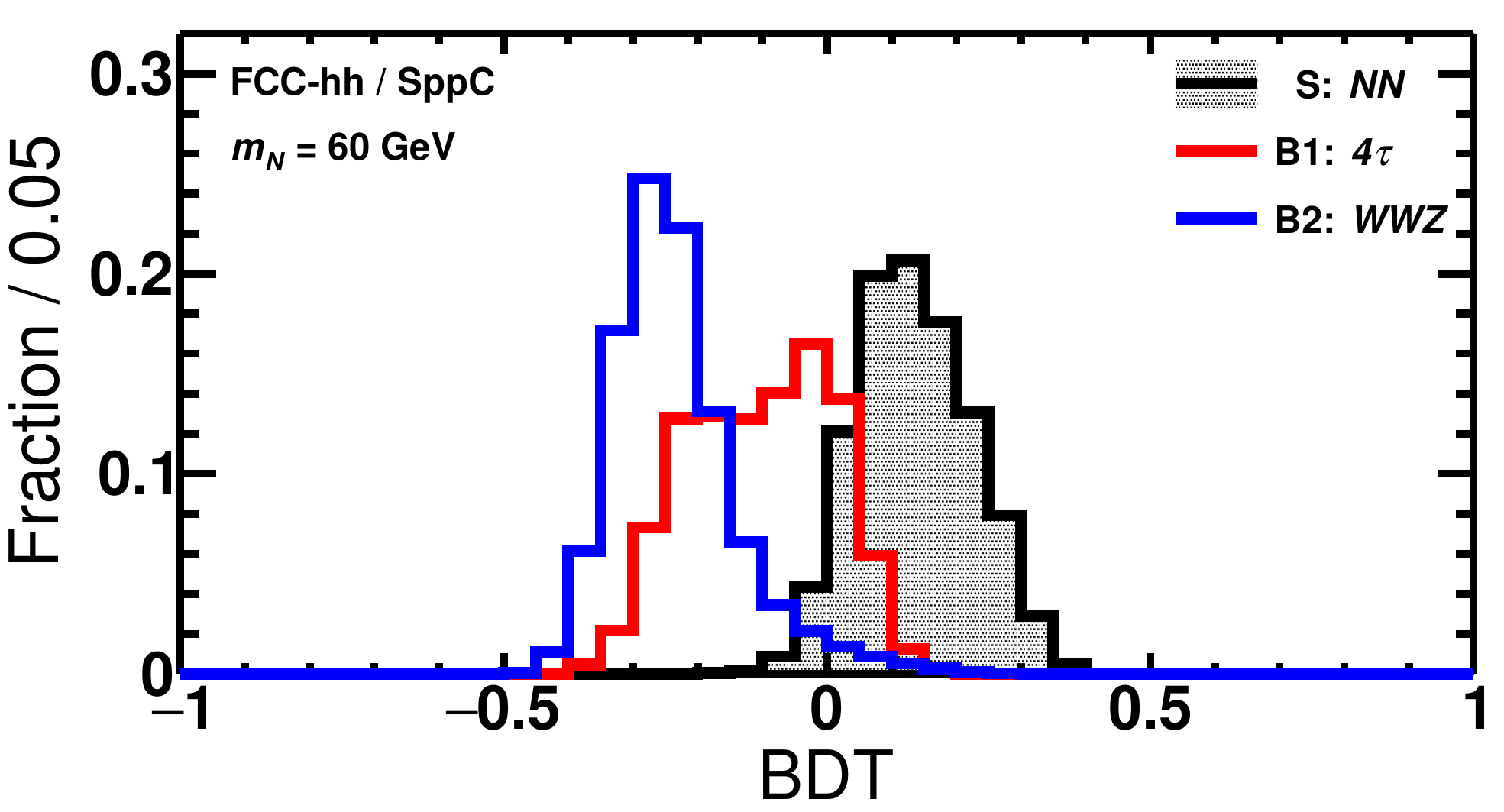}   
\caption{The BDT distributions of the same processes as in Fig.~\ref{fig:BDT14TeV}, but at the FCC-hh/SppC.}
\label{fig:BDT100TeV}
\end{figure}

\section{Selection efficiency table}
\label{appendix:efficiencies}

In Table~\ref{tab:allEfficiencies}, we show selection efficiencies for both signal $NN$ and background processes at the HL-LHC and FCC-hh/SppC for different heavy neutrino masses.
The total selection efficiency after both pre-selection and BDT criteria is the product of two efficiencies in the same column.
The numbers of events for background after all selections can be calculated by multiplying the initial numbers in the Table~\ref{tab:cutFlow14TeV} by the total cut efficiency, while the number of events for signal can be calculated from Eq.~(\ref{eq:xsec}) by substituting the total cut efficiency for $NN$ as $A_{\rm eff}^{(1)}$.

\begin{table}[H]
\centering 

\begin{tabular}{cccccc}

\hline
\hline
$m_N$ &  $\sqrt{s}$ & selection  & $NN$                       & $4\tau$                    & $WWZ$ \\
\hline
\multirow{4}{*}{10 GeV} & \multirow{2}{*}{14 TeV}    & pre-selection & $1.10\mltp10^{-1}$&$2.41\mltp10^{-5}$&$1.33\mltp10^{-3}$\\
                                         &                                            & BDT$>$0.428  &$9.52\mltp10^{-1}$&$2.86\mltp10^{-3}$&$1.50\mltp10^{-4}$\\
                                         & \multirow{2}{*}{100 TeV} & pre-selection&$2.26\mltp10^{-1}$&$1.01\mltp10^{-4}$&$3.25\mltp10^{-3}$\\
                                         &                                            & BDT$>$0.311 &$8.97\mltp10^{-1}$&$9.46\mltp10^{-4}$&$1.57\mltp10^{-5}$\\
\hline
\multirow{4}{*}{20 GeV} & \multirow{2}{*}{14 TeV}    &pre-selection&$1.08\mltp10^{-1}$&$2.41\mltp10^{-5}$&$1.33\mltp10^{-3}$\\
                                         &                                            &BDT$>$0.267  &$8.60\mltp10^{-1}$&$9.43\mltp10^{-3}$&$6.87\mltp10^{-4}$\\
                                         & \multirow{2}{*}{100 TeV} & pre-selection&$2.32\mltp10^{-1}$&$1.01\mltp10^{-4}$&$3.25\mltp10^{-3}$\\
                                         &                                            & BDT$>$0.170  &$8.69\mltp10^{-1}$&$6.30\mltp10^{-3}$&$4.25\mltp10^{-4}$\\
\hline
\multirow{4}{*}{30 GeV} & \multirow{2}{*}{14 TeV}    & pre-selection&$1.12\mltp10^{-1}$&$2.41\mltp10^{-5}$&$1.33\mltp10^{-3}$\\
                                         &                                            & BDT$>$0.177  &$6.82\mltp10^{-1}$&$1.54\mltp10^{-2}$&$1.96\mltp10^{-3}$\\
                                         & \multirow{2}{*}{100 TeV} & pre-selection&$2.56\mltp10^{-1}$&$1.01\mltp10^{-4}$&$3.25\mltp10^{-3}$\\
                                         &                                            &BDT$>$ 0.150 &$5.56\mltp10^{-1}$&$5.25\mltp10^{-3}$&$1.22\mltp10^{-3}$\\
\hline
\multirow{4}{*}{40 GeV} & \multirow{2}{*}{14 TeV}    & pre-selection&$1.35\mltp10^{-1}$&$2.41\mltp10^{-5}$&$1.33\mltp10^{-3}$\\
                                         &                                            &BDT$>$0.122  &$6.58\mltp10^{-1}$&$3.82\mltp10^{-2}$&$9.69\mltp10^{-3}$\\
                                         & \multirow{2}{*}{100 TeV} & pre-selection&$3.07\mltp10^{-1}$&$1.01\mltp10^{-4}$&$3.25\mltp10^{-3}$\\
                                         &                                            & BDT$>$0.131  &$4.35\mltp10^{-1}$&$5.56\mltp10^{-3}$&$3.59\mltp10^{-3}$\\
\hline
\multirow{4}{*}{50 GeV} & \multirow{2}{*}{14 TeV}    & pre-selection&$1.56\mltp10^{-1}$&$2.41\mltp10^{-5}$&$1.33\mltp10^{-3}$\\
                                         &                                            & BDT$>$0.128  &$5.34\mltp10^{-1}$&$2.46\mltp10^{-2}$&$1.13\mltp10^{-2}$\\
                                         & \multirow{2}{*}{100 TeV} & pre-selection&$3.62\mltp10^{-1}$&$1.01\mltp10^{-4}$&$3.25\mltp10^{-3}$\\
                                         &                                            & BDT$>$0.117  &$4.71\mltp10^{-1}$&$9.42\mltp10^{-3}$&$7.20\mltp10^{-3}$\\
\hline
\multirow{4}{*}{60 GeV} & \multirow{2}{*}{14 TeV}    & pre-selection&$1.77\mltp10^{-1}$&$2.41\mltp10^{-5}$&$1.33\mltp10^{-3}$\\
                                         &                                            &BDT$>$0.138  &$5.88\mltp10^{-1}$&$1.79\mltp10^{-2}$&$1.16\mltp10^{-2}$\\
                                         & \multirow{2}{*}{100 TeV} & pre-selection&$4.03\mltp10^{-1}$&$1.01\mltp10^{-4}$&$3.25\mltp10^{-3}$\\
                                         &                                            & BDT$>$0.131  &$4.95\mltp10^{-1}$&$5.20\mltp10^{-3}$&$5.06\mltp10^{-3}$\\

\hline
\hline                                         

\end{tabular}
\caption{
Selection efficiencies for the signal $NN$ and background processes of $4\tau$ and $WWZ$ at the HL-LHC (with center-of-mass energy $\sqrt{s}=14$ TeV) and FCC-hh/SppC (with $\sqrt{s}=100$ TeV) for $m_N = $ 10, 20, 30, 40, 50, 60 GeV, respectively. Here $h_2$ is assumed to be heavy and only $h_1$ decay contributes to the signal $NN$.
}
\label{tab:allEfficiencies}
\end{table}

\acknowledgments

We thank Monica D’Onofrio and Yusheng Wu for helpful discussions. 
Y.G. is supported by the IHEP, CAS under the CEPC theory grant (2019-2020) and partially under grant no.~Y7515560U1. 
M.J. by the National Natural Science Foundation of China under grant no.~11475191.
K.W. by the Excellent Young Talents Program of the Wuhan University of Technology, the National Natural Science Foundation of China under grant no.~11905162, and the CEPC theory grant (2019-2020) of IHEP, CAS.



\end{document}